\documentclass{article}
\pdfoutput=1
\usepackage{jheppub}

\usepackage{graphicx, subcaption}% Include figure files
\usepackage{dcolumn}% Align table columns on decimal point
\usepackage{bm}% bold math
\usepackage{xcolor}
\usepackage{amsmath}
\usepackage{multirow}
\usepackage{float}
\usepackage{hyperref}
\usepackage{tabularx}
\usepackage[numbers]{natbib}
\DeclareRobustCommand{\Sec}[1]{Sec.~\ref{#1}}

\DeclareRobustCommand{\App}[1]{App.~\ref{#1}}
\DeclareRobustCommand{\Tab}[1]{Table~\ref{#1}}

\DeclareRobustCommand{\Fig}[1]{Fig.~\ref{#1}}
\DeclareRobustCommand{\Figs}[2]{Figs.~\ref{#1} and \ref{#2}}

\DeclareRobustCommand{\Feta}[0]{\textsc{FETA}}
\DeclareRobustCommand{\Cathode}{\textsc{CATHODE}}
\DeclareRobustCommand{\Curtains}{\textsc{CURTAINs}}
\DeclareRobustCommand{\Salad}{\textsc{SALAD}}

\title{The Interplay of Machine Learning--based \\ Resonant Anomaly Detection Methods}

\affiliation[a]{Département de physique nucléaire et corpusculaire, Université de Genève, 1211 Genève, Switzerland}
\affiliation[b]{Institut f\"{u}r Experimentalphysik, Universit\"{a}t Hamburg, 22761 Hamburg, Germany}
\affiliation[c]{Institut f\"ur Theoretische Physik, Universit\"at Heidelberg, 69120 Heidelberg, Germany}
\affiliation[d]{Department of Physics, University of California, Berkeley, CA 94720, USA}
\affiliation[e]{Physics Division, Lawrence Berkeley National Laboratory, Berkeley, CA 94720, USA}
\affiliation[f]{Berkeley Institute for Data Science, University of California, Berkeley, CA 94720, USA}
\affiliation[g]{NHETC, Dept. of Physics and Astronomy, Rutgers University, Piscataway, NJ 08854, USA}

\author[a]{Tobias Golling,}
\author[b]{Gregor Kasieczka,}
\author[c]{Claudius Krause,}
\author[d,e]{Radha Mastandrea,}
\author[e,f]{Benjamin Nachman,}
\author[a]{John Andrew Raine,}
\author[a]{Debajyoti Sengupta,}
\author[g]{David Shih,}
\author[b]{and Manuel Sommerhalder}

\emailAdd{tobias.golling@unige.ch}
\emailAdd{gregor.kasieczka@uni-hamburg.de}
\emailAdd{claudius.krause@thphys.uni-heidelberg.de}
\emailAdd{rmastand@berkeley.edu}
\emailAdd{bpnachman@lbl.gov}
\emailAdd{john.raine@unige.ch}
\emailAdd{debajyoti.sengupta@unige.ch}
\emailAdd{shih@physics.rutgers.edu}
\emailAdd{manuel.sommerhalder@uni-hamburg.de}

\abstract{
Machine learning--based anomaly detection (AD) methods are promising tools for extending the coverage of searches for physics beyond the Standard Model (BSM).  One class of AD methods that has received significant attention is resonant anomaly detection, where the BSM physics is assumed to be localized in at least one known variable.  While there have been many methods proposed to identify such a BSM signal that make use of simulated or detected data in different ways, there has not yet been a study of the methods' complementarity.  To this end, we address two questions.  First, in the absence of any signal, do different methods pick the same events as signal-like?  If not, then we can significantly reduce the false-positive rate by comparing different methods on the same dataset.  Second, if there is a signal, are different methods fully correlated?  Even if their maximum performance is the same, since we do not know how much signal is present, it may be beneficial to combine approaches.  Using the Large Hadron Collider (LHC) Olympics dataset, we provide quantitative answers to these questions.  We find that there are significant gains possible by combining multiple methods, which will strengthen the search program at the LHC and beyond.}

\begin{document}

\maketitle

\section{Introduction}

Since the observation of the Higgs Boson in 2012 at the Large Hadron Collider (LHC)~\citep{Aad:2012tfa,Chatrchyan:2012ufa}, no new fundamental particle has been observed. This is not for lack of effort: theoretical models involving supersymmetric particles, dark matter candidates, or heavy matter generations abound, informing past, current, and planned analyses at the LHC~\citep{atlasexoticstwiki,atlassusytwiki,atlashdbspublictwiki,cmsexoticstwiki,cmssusytwiki,cmsb2gtwiki,lhcbtwiki}.  Given that such past searches for specific alternatives to the Standard Model (SM) have been unsuccessful, there has been a push to run broader, model-agnostic searches for new physics in parallel. In particular, machine learning (ML) has enabled many new search strategies~\citep{Karagiorgi:2021ngt,Kasieczka:2021xcg,Aarrestad:2021oeb}.

One of the most popular and well-motivated search strategies for evidence of physics beyond the Standard Model is \textit{resonant anomaly detection}. In such investigations, the new physics signal is expected to take the form of a new particle, i.e.~a resonance with respect to a mass-like event variable. The search strategy then involves looking for a localized excess of these new physics events with respect to the SM background.

There now exist many ML methods for resonant anomaly detection (AD)\footnote{We are not counting generic AD methods applied to the resonant case, see e.g. the recent ATLAS results~\citep{ATLAS:2023azi,ATLAS-CONF-2023-022} and method papers they cite.} with comparable sensitivities \citep{Collins:2018epr,Collins:2019jip,Andreassen:2020nkr,Nachman:2020lpy, Benkendorfer:2020gek, Stein:2020rou, Amram_2021, Hallin:2021wme, Kamenik:2022qxs, Hallin:2022eoq, Chen:2022suv, Golling:2022nkl,sengupta2023curtains}, some of which have also been applied to data~\citep{ATLAS:2020iwa,Shih:2021kbt,Shih:2023jfv,Pettee:2023zra}. These methods have largely been developed independently of each other, with different strengths and weaknesses. However, there has not yet been a thorough study of the complementarity of these techniques. In particular, we want to ask the questions: can we improve signal sensitivity by combining these methods? Can we improve robustness in the background-only case by combination? Do these different methods classify the same things as ``signal-like" for background and signal events?

In this paper, we evaluate a selection of these resonant AD methods on equal footing, using an identical methodological setup for each. In \Sec{sec:methodology}, we provide a more detailed background of the resonant AD procedure and introduce the four detection methods that we will consider in this paper. In \Sec{sec:contrasting}, we consider how similar the detection methods are to each other, gauging whether different methods pick up distinct components of the phase space of resonant anomalies. In \Sec{sec:combining}, we combine the four sample generation methods with the goal of increasing sensitivity for a resonant AD task. We conclude in \Sec{sec:conclusions}, suggesting avenues for further exploration. 

As a word of caution: this study is not meant to be an exhaustive summary for machine learning-enhanced anomaly detection across all signals and setups. For illustrative purposes, we focus on one well-studied signal model and signal region and compare our findings with the existing literature. Practitioners should examine different methods in their own region of phase space.

\section{Methodology}
\label{sec:methodology}

\subsection{Overview of resonant anomaly detection}

\begin{figure}
    \centering
    \includegraphics[width = 0.5\textwidth]{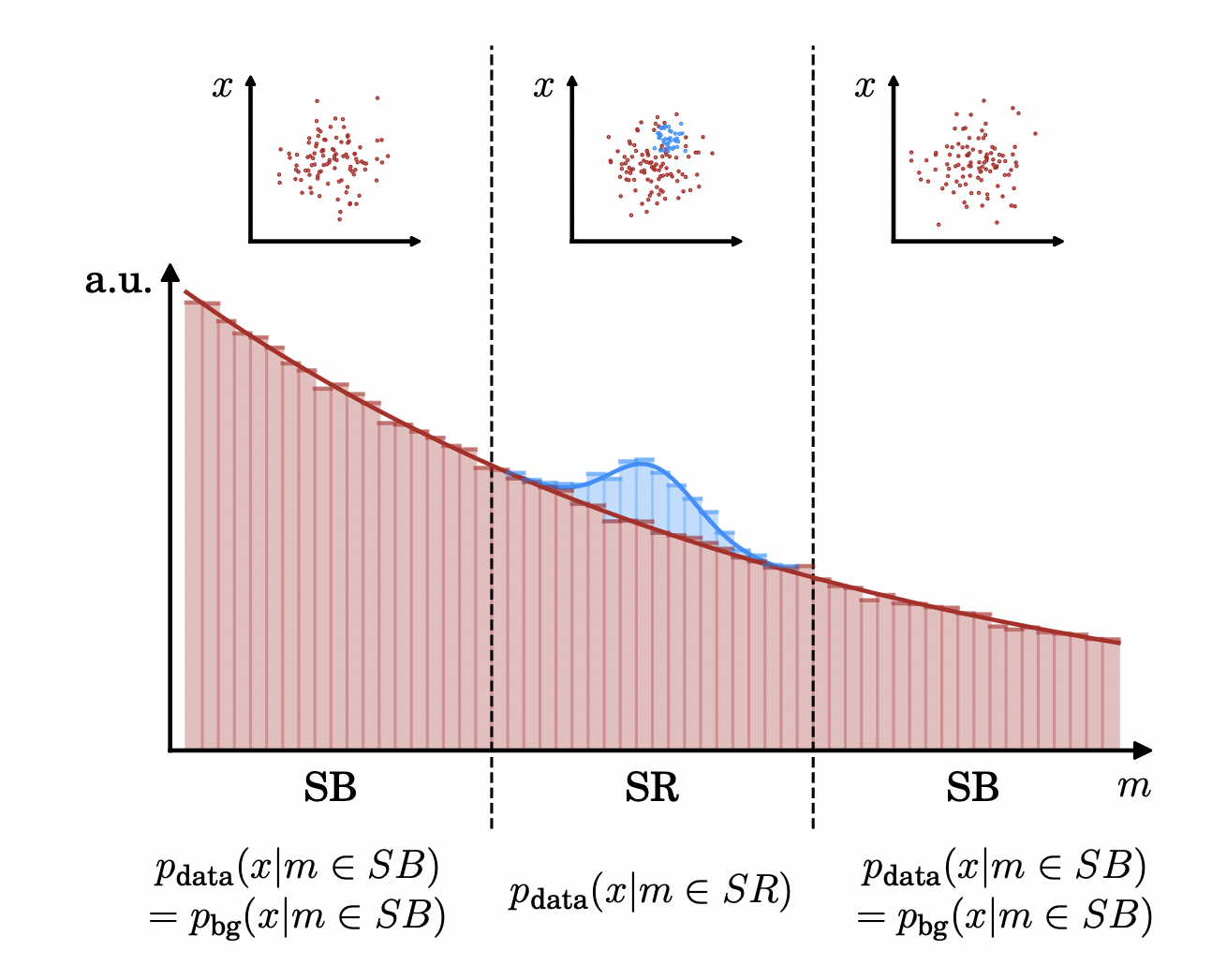}
    \caption{A schematic of the resonant anomaly detection motivation. The goal is to observe an excess of signal (blue) events above a background (red) process. The signal is localized in $m$ to a signal region (SR), and a model for background can be derived from data in the sidebands (SB) regions. Typically, the signal-background discrimination task makes use of features other than $m$. Figure is taken from \citep{Hallin:2021wme}.}
    \label{fig:bump_hunt}
\end{figure}

The goal of resonant AD (illustrated schematically in \Fig{fig:bump_hunt}) is to find an excess of beyond-the-Standard Model (BSM) events that are localized in some event variable $m$ (usually, a mass-like feature). The BSM signal thus corresponds to a new particle with a nonzero $m$, expected to be reconstructed within a \textit{signal region}, SR, defined as an interval in $m$. In particular, the search makes use of a set of other (i.e. non-$m$) features in order to elevate the sensitivity above that of a standard bump hunt. Importantly, the excess events must be observed with respect to a SM background, but this background is nontrivial to construct: using out-of-the-box simulated data is not ideal given the numerous necessary approximations made for the hard-process, showering, and detector simulation steps; using actual data from outside of the signal region (or in \textit{sideband regions}, SB) requires the analysis to only use event features that are statistically independent from the mass variable~\citep{Collins:2018epr,Collins:2019jip,Benkendorfer:2020gek}.

An alternative strategy is to construct a set of \textit{synthetic SM samples}, or events that are representative of the SM background process in the same mass space as the BSM events. A binary classifier  trained to discriminate the synthetic samples from detected data is then the optimal classifier for discriminating SM background from the new physics (see Ref.~\citep{Metodiev:2017vrx}), so long as the synthetic samples are indeed a faithful representation of actual SM (i.e. not containing any events derived from a resonant anomaly) events in the probed mass range.

In recent years, there has been much work done on developing procedures to construct such synthetic SM samples. While there now exist many varied methods for sample construction, the vast majority\footnote{The vast majority, but not all. For examples of methods that cannot be so neatly classified on these two characters, see \citep{Benkendorfer:2020gek, Amram_2021, Choi:2020bnf, CMS:2022uga}.} of them can be characterized based on two properties of their construction. First, generation methods can be \textit{data-exclusive} or \textit{simulation-assisted}: data-exclusive methods generate synthetic samples by making use of collider data from the SB mass regions, where the data are far enough from any BSM signal to be treated as representative of SM background; simulation-assisted methods will also use an auxiliary dataset of simulated background-only collisions. Second, generation methods exploit machine learning techniques through either \textit{likelihood(-ratio) learning} or \textit{feature morphing}: methods can either learn the likelihood(-ratio) of an SM-only dataset (this can be either from the auxiliary simulated dataset or the SB data) and interpolate this likelihood into the SR; alternatively, methods can morph features from said background-only regions into the SR.

In this paper, we will consider the four methods shown in \Tab{tab:methods}, which span this ``character space" of methods for resonant AD. 

\begin{table}[h]
%\footnotesize
    \centering
    \begin{tabular}{|c|c|c|}
    \hline
      & Simulation-assisted  & Data-exclusive  \\
      \hline
      Likelihood learning & \textsc{Salad} \citep{Andreassen:2020nkr}  &  \textsc{(La)Cathode}  \citep{Hallin:2021wme, Hallin:2022eoq} \\
      \hline
      Feature morphing   & \textsc{Feta} \citep{Golling:2022nkl} & \textsc{Curtains} \citep{Raine:2022hht,sengupta2023curtains} \\
      \hline
    \end{tabular}
    \caption{Many methods for constructing Standard Model background templates for resonant anomaly detection can be classified on two axes: on the horizontal, usage of an auxiliary dataset (simulation); on the vertical, how non-signal region Standard Model background processes are morphed into signal region Standard Model template samples.}
    \label{tab:methods}
\end{table}

We provide a brief summary of the four methods considered here.

\begin{itemize}
    \item \Salad{}: Simulation Assisted Likelihood-free Anomaly Detection \citep{Andreassen:2020nkr} trains a binary classifier to discriminate simulated SM events from detected SM events in the SB (background-only) region, then uses the classifier to reweight simulated SM events in the SR. These reweighted events comprise the synthetic SM samples.
     \item \Cathode{}: Classifying Anomalies THrough Outer Density Estimation \citep{Hallin:2021wme} trains a normalizing flow--based probability density estimator to model detected data in SB, then interpolates the distribution into the SR. A set of events drawn from the interpolated distribution comprises the synthetic SM samples.
    \item \Curtains{}: Constructing Unobserved Regions by Transforming Adjacent INtervals \citep{Raine:2022hht,sengupta2023curtains} trains a normalizing flow--based transport function to morph detected data between low- and high-mass SB, then applies the flow to map from SB into the SR. These morphed samples comprise the synthetic SM samples.
    \item \Feta{}: Flow-Enhanced Transportation for Anomaly detection \citep{Golling:2022nkl} trains a normalizing flow--based transport function to morph SM simulation in SB to detected data in background-dominated SB, then applies the model to SM simulation in the SR. These morphed samples comprise the synthetic SM samples. 
   
\end{itemize}

\begin{figure}
    \centering
    \includegraphics[width=0.5\textwidth]{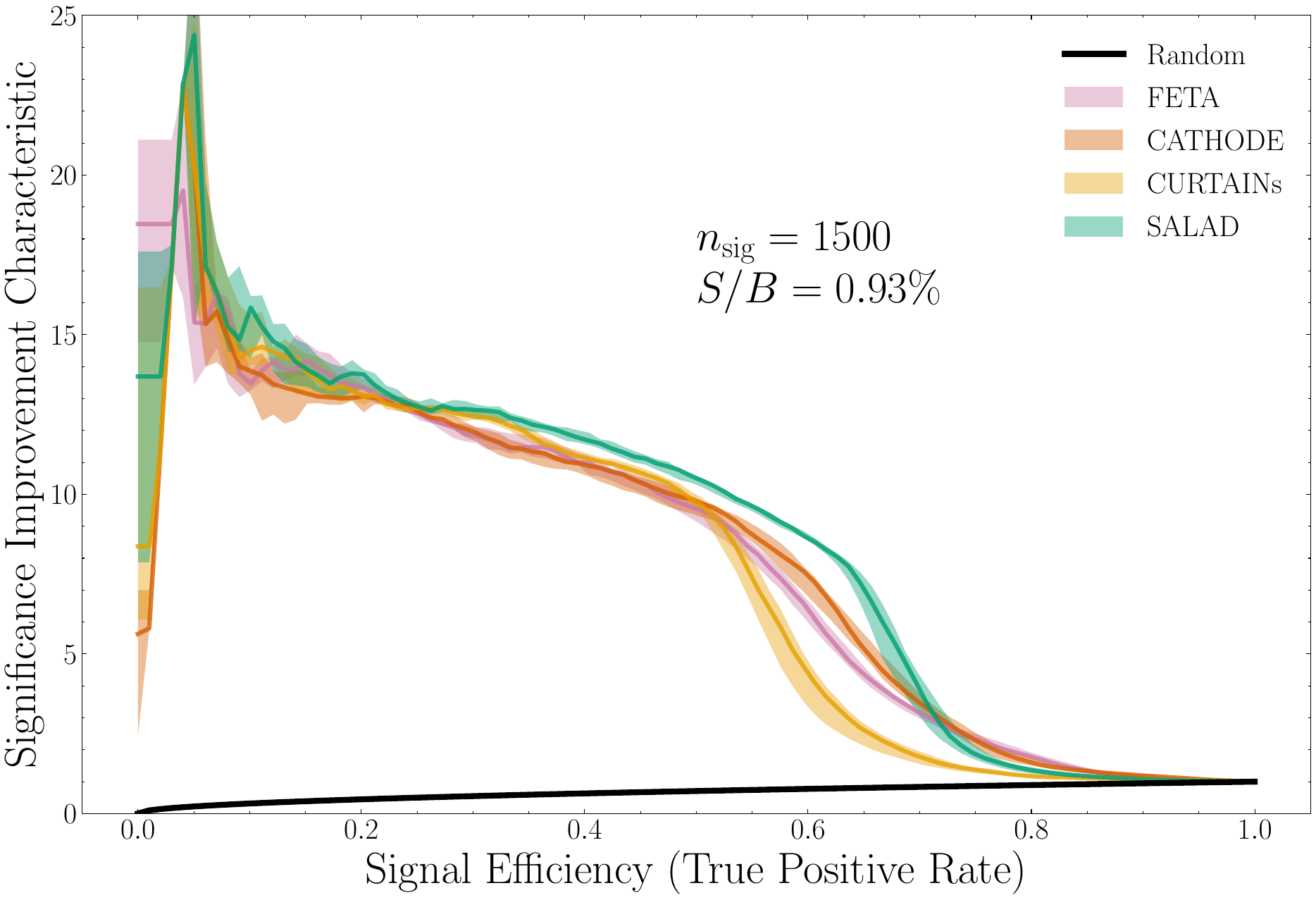}
    \caption{Significance Improvement Characteristic for a binary classifier trained to discriminate each synthetic background generation method's samples from signal-contaminated data. Below this signal injection ($n_\mathrm{sig} = 1500$), methods perform less equally. For readability, we withhold a description of the classifier architectures and ensembling choices until \Sec{sec:classifier_architecture}.}
    \label{fig:sic_1500}
\end{figure}

Our goal is then to explore how the synthetic SM samples generated by each of these  methods perform in resonant AD tasks, focusing on their relative performances in addition to their absolute performances. In fact, all four methods are comparable at picking up on signal contaminations of $\sim0.93\%$ and above: in \Fig{fig:sic_1500}, we plot the significance improvement characteristic (SIC) as a function of the signal efficiency. Broadly speaking, the SIC corresponds to the multiplicative factor by which a signal significance would improve by making a well-motivated cut on the data; a classifier that is ideally suited to discriminating signal from background would have a high SIC at all signal efficiencies.

\subsection{Dataset}

We use the LHC Olympics 2020 R\&D dataset \citep{LHCOlympics,Kasieczka:2021xcg}, which consists of 1,000,000 background events comprised of QCD dijet production, together with 100,000 signal events from a $Z'$ resonance at 3.5 TeV,
decaying to scalars $X$ and $Y$ at 500~GeV and 100~GeV respectively, which then each decay to quark pairs. Since the $X$ and $Y$ scalars are highly boosted, their decay products are highly collimated and form large-radius jets. For the main resonant feature, we use the dijet invariant mass ($m=m_{JJ}$) which should reconstruct the $Z'$ mass for the signal.
Events are required to have at least one large-radius anti-$k_T$~\citep{Cacciari:2011ma,Cacciari:2008gp} jet ($R=1$) with a $p_T$ threshold of $1.2$ TeV. Each event contains up to 700 particles with three degrees of freedom $p_T$, $\eta$, $\phi$. The events are generated  with \textsc{Pythia}~8.219~\citep{Sjostrand:2006za,Sjostrand:2014zea} and   \textsc{Delphes}~3.4.1~\citep{deFavereau:2013fsa}. Also included in the LHC Olympics dataset is a set of 1,000,000 \textsc{Herwig}++\citep{B_hr_2008} QCD dijet events generated using the same tunes, which is used for the simulation-assisted approaches (\Salad{} and \Feta{}) as the auxiliary ``simulated" data (SIM). We therefore denote the \textsc{Pythia} events as ``data" (DAT).

In addition to the LHC Olympics dataset, we make use of two auxiliary sets of QCD dijet events generated using the same tunes as the LHC Olympics dataset, but only in the $m_{JJ}$ region [3.3, 3.7] TeV (our chosen signal region). The first set, consisting of 1,000,000   \textsc{Herwig}++-based events, is used to generate additional SALAD samples. The second set, consisting of 320k \textsc{Pythia}-based events, is used for testing the synthetic samples.

%Our BSM signal of interest comes from the process $Z'\rightarrow X(\rightarrow q\overline{q})Y(\rightarrow q\overline{q})$, which actually contains three new resonances $Z'$ ($3.5$~TeV), $X$ (500~GeV), and $Y$ (100~GeV). (We do not explicitly study the smaller two resonances in this paper.)

%with respect to a SM QCD background. 

%dijet events and 100,000 signal dijet events

\begin{figure}
    \centering
    \includegraphics[width = .7\linewidth]{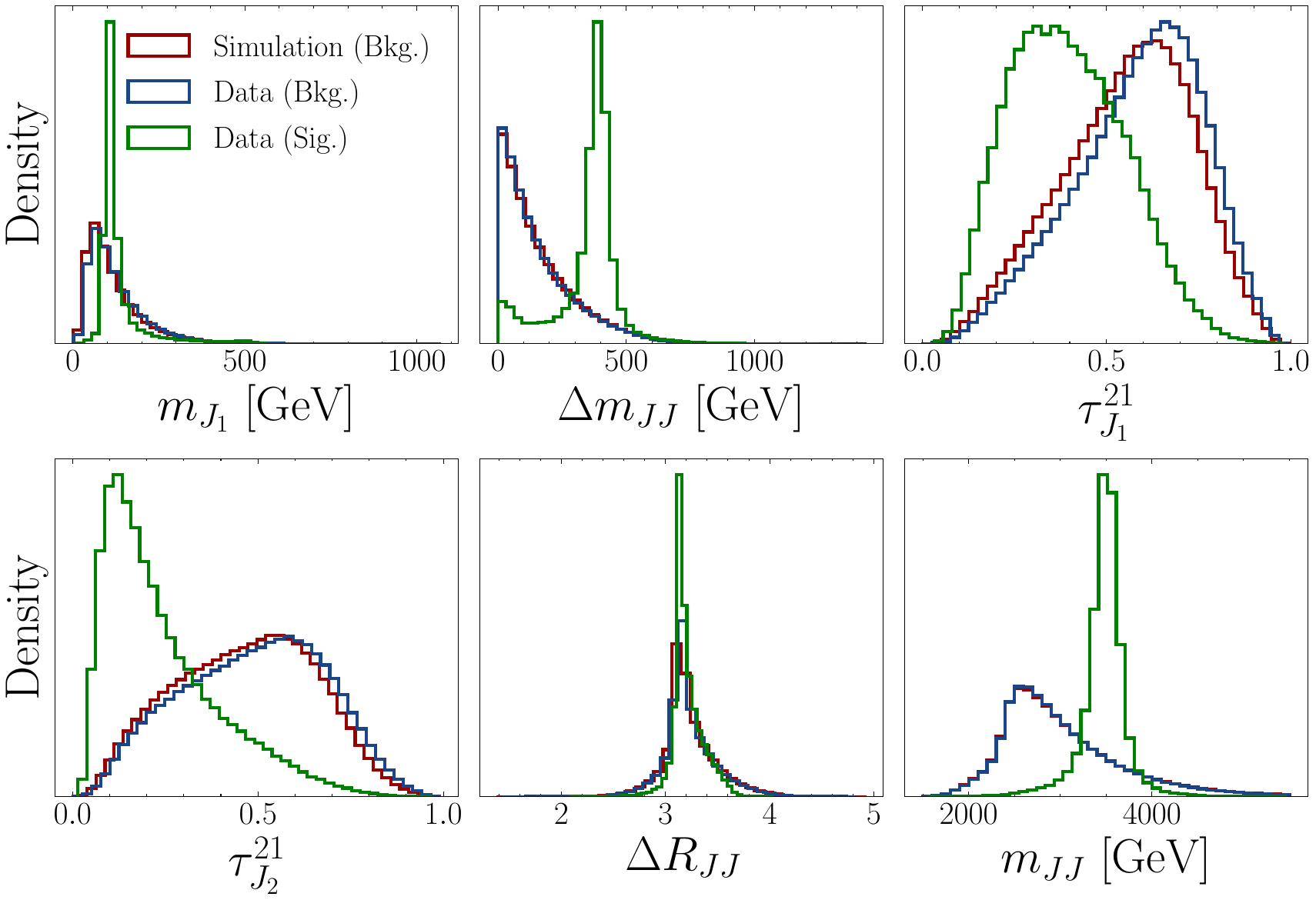}
    \caption{The 5-dimensional feature space of dijet collision events used in this resonant AD study. We also show a 6th feature, a mass-like event variable that is used to define a signal region (SR) and sidebands regions (SB).}
    \label{fig:features}
\end{figure}

We choose a feature space of six dijet observables $m_{J_1}$, $\Delta m_{JJ}$, $\tau^{21}_{J_1}$, $\tau^{21}_{J_2}$, $\Delta R_{JJ}$, and $m_{JJ}$ (see \Figs{fig:features}{fig:background_validation}). This last feature is our mass-like event variable, which we use to define a SR spanning [3.3, 3.7] TeV. For our sidebands, we use the full amount of available data, down to 1.5 TeV and up to 5.5 TeV. 

Synthetic samples are generated using the procedure outlined in each method's respective paper. However, in an attempt to equalize the quality of the samples as much as possible, all methods are given the same training and validation sets of simulation and data, which is an 80-20 split of the full sidebands data, i.e. all data in the range ([1.5, 5.5] $\setminus$ [3.3, 3.7]) TeV. These training and validation event counts, as well as the number of synthetic SM samples generated for each method, are shown in \Tab{tab:synthetic_samples_counts}. 

The \Cathode{}, \Curtains{}, and \Feta{} methods all involve training a generative model (i.e. a normalizing flow) to learn an underlying probability distribution. This allows us to reduce statistical uncertainties of the synthetic SM samples by \textit{oversampling} from the generative models. For \Cathode{}, we can sample from the normalizing flow that has been interpolated into the SR as many times as we want; for \Curtains{} (or \Feta{}), we can similarly oversample from the normalizing flows that learn the densities of SB data (or SR simulation). In this study, we use the oversampling factor (also shown in \Tab{tab:synthetic_samples_counts}) that was shown to saturate each model's performance in its respective paper. 

\begin{table}[]
    \centering
    \begin{tabular}{|c|c|c|c|c|}
    \hline
        Method & Training data & Validation data & \# samples & Oversampling \\
        \hline
         \hline
          \Salad{} & 793k SIM, 696k DAT & 198K SIM, 174K DAT & 1,045k & N/A \\
         \hline
          \Cathode{} & 696k DAT & 174K DAT & 400k & 3 \\
         \hline
        \Curtains & 373k DAT & 93k DAT & 1,887k & 4\\
         \hline
        \Feta{} & 793k SIM, 696k DAT & 198K SIM, 174K DAT & 732k & 6\\
         \hline
    \end{tabular}
    \caption{Numerical breakdown of the events used to construct the given number of synthetic SM samples (in the SR) for the four machine learning methods considered in this report. The \Curtains{} method uses a slightly narrower SB region of [2.7, 4.5] TeV to avoid transforming events across the $m_{JJ}$ turn-on region border. The \Salad{} samples are generated by applying the learned weights to an additional, much larger set of \textsc{Herwig++} simulated SM events not contained in the LHC Olympics dataset. Note that \Cathode{} and \Curtains{} are data-exclusive (i.e. fully data-driven), using only the the ``detected" (DAT) \textsc{Pythia} set, while \Salad{} and \Feta{} require an auxiliary ``simulated" (SIM) \textsc{Herwig++} set. }
    \label{tab:synthetic_samples_counts}
\end{table}

\subsection{Classifier architecture}
\label{sec:classifier_architecture}

As stated earlier, a set of synthetic SM background samples can be used for a resonant AD task by training a binary classifier to discriminate the samples from a set of detected SR data. If the SR data contains a nonzero percentage of BSM events, then the binary classifier should be able to pick up on this difference. In fact, that classifier (in the limit of infinite data and arbitrarily flexible training / architecture) is the optimal one for distinguishing SM background from the BSM signal~\citep{Metodiev:2017vrx}.

Concretely, the discrimination task in this paper is between a given method's synthetic SM samples and a set of 121k SR DAT background + $n_\mathrm{sig}$ SR DAT signal ($Z'$) events, where $n_\mathrm{sig}$ is a controllable parameter, the number of injected signal events in the SR.

To evaluate the similarity between two datasets of events, we use a binary classifier network consisting of 3 linear layers of size (64, 32, 1). The network uses rectified linear unit (\textsc{ReLU}) activations and is optimized using \textsc{Adam} \citep{adam} on the Binary Cross Entropy loss. We train using a 5-fold cross validation scheme, keeping the network with the best validation loss. Each fold is trained with a batch size of 128\footnote{Interestingly, we found that the choice of batch size significantly affected the classifier, with larger batch sizes leading to sizeable drop-offs in performance.} and a learning rate of $10^{-3}$ for up to 100 epochs with a patience of 5 epochs. All hyperparameters were optimized via manual tuning to give the best possible significance improvement characteristic curves (introduced formally in \Sec{sec:combining})\footnote{Note that since the binary classifier is trained on signal region data, the network is no longer signal-agnostic.}.  All errorbars and errorbands in plots come from retraining the binary classifier with different random seeds. Note that we do not vary the random seeds for training the architectures that generate each set of synthetic samples, but we find these effects to be small in \App{app:robust}.

All figures and summary statistics are generated by evaluating the trained binary classifier networks on a ``standard test set" of 20,000 signal and 320,000 background dijet events, which are not used at any time during the training procedure. Using a larger number of background events for evaluation allows for smooth summary statistics at low signal efficiencies, which is expected to be the relevant region for resonant AD. Unless explicitly stated otherwise, all plots are score-averaged for each method, i.e. the plots are derived from the average of classifier scores over 10 independent runs.

\section{Contrasting the synthetic SM samples}
\label{sec:contrasting}

\subsection{Background-only case}
\label{sec:bkg_only}

\begin{figure}
    \centering
\includegraphics[width=\linewidth]{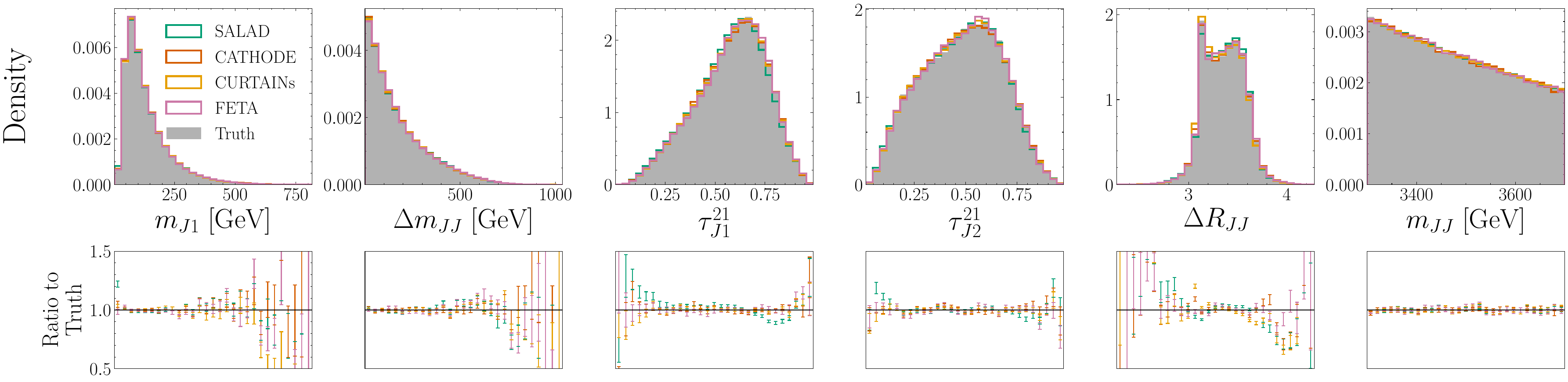}
    \caption{Distributions of synthetic SM backgrounds generated by each method compared to data with $n_\mathrm{sig} = 0$.}
    \label{fig:background_validation}
\end{figure}

As a first test of the synthetic SM samples created by each of the four generation methods, we compare their distributions to background-only SM data, i.e. with a signal injection $n_\mathrm{sig} = 0$. These distributions are shown in \Fig{fig:background_validation}. We see that at this level all four methods reproduce the true distribution well. In particular, the ratios of marginals for all methods are all close to 1, which indicates that any differences between the sample generation methods and truth cannot be ascribed to a single observable. As a quantitative assessment of the marginal distributions, in \Tab{tab:ks_test_stat}, we provide the Kolmogorov-Smirnov (KS) test statistics for the marginal distributions between each method and the truth. The KS test statistic is defined as the supremum of the differences between two distributions' empirical cumulative distribution functions, and can therefore provide a gauge of how different two distributions are. 

\begin{table}[]
    \centering
    \begin{tabular}{|c|c|c|c|c|c|c|}
    \hline
        Method & $m_{{J1}}$& $\Delta m_{{JJ}}$ & $\tau_{{J1}}^{{21}}$& $\tau_{{J2}}^{{21}}$&$\Delta R_{{JJ}}$ & $m_{{JJ}}$ \\
        \hline
         \hline
          \Salad{} & 0.00775 & 0.00501 & 0.02229 & 0.00610 & 0.01205 & 0.00215  \\
         \hline
          \Cathode{} & 0.00405 & 0.00450 & 0.00597 & 0.00534 & 0.00755 & 0.00228\\
         \hline
        \Curtains & 0.00325 & 0.00255 & 0.00238 & 0.00214 & 0.02122 & 0.00353\\
         \hline
        \Feta{} & 0.00605 & 0.00352 & 0.00588 & 0.00536 & 0.00725 & 0.00386\\
         \hline
    \end{tabular}
    \caption{Kolmogorov-Smirnov test statistics between each method's marginal distribution and the truth's marginal distribution. Larger test statistics indicate a greater difference between two distributions. as gauged by the maximum difference in the empirical cumulative distribution functions. }
    \label{tab:ks_test_stat}
\end{table}

As a next test of the synthetic SM samples created by each of the four generation methods, we analyze classifiers trained to discriminate synthetic SM background from background-only SM data, i.e. with a signal injection $n_\mathrm{sig} = 0$. Given that there is no BSM signal present in the training data, differences in classifier performances come down to the ``nature" of the synthetic SM samples for each method. All of the methods are given the same training and validation data sets, so the statistical fluctuations from the input data should be correlated between methods.  Further, all of the binary classifiers are evaluated with the same random seed, so the network initializations should be identical in that respect. However, there are differences stemming from the initialization of the neural networks for each \textit{method}, as well as from the differences of the methods themselves. These differences might be expected to decorrelate the classifier scores.

\begin{figure}
    \centering
    \includegraphics[width = .5\textwidth]{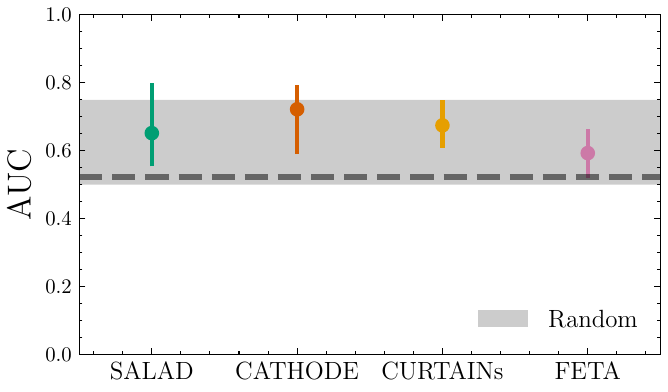}
    \caption{Receiver operating characteristic area-under-the-curves (AUC) for binary classifiers trained to discriminate each method's synthetic samples from data with $n_\mathrm{sig} = 0$. The table summarizes 100 classifier runs with different random seeds, with errorbars showing a 68-percentile spread. Also plotted is the spread of 100 random classifiers, with the thick dashed line showing the median of those runs. No score averaging has been done for this plot. To see the AUC spreads corresponding to the combination of samples, see \Fig{fig:aucs_w_combined}.}
    \label{fig:aucs}
\end{figure}

We provide the receiver operating characteristic area-under-the-curves (AUC) for such classifiers in \Fig{fig:aucs}. Also plotted is the AUC spread derived from training a classifier to discriminate truth from truth, which represents the spread of a random classifier given the set of different network initializations and the fact that the network is not infinitely powerful. The ROC spread of each individual method is consistent with that of the spread of a random classifier, again providing evidence that the nature of the synthetic samples is truth- (SM-) like.

\begin{figure}
    \centering
\includegraphics[width=.6\linewidth]{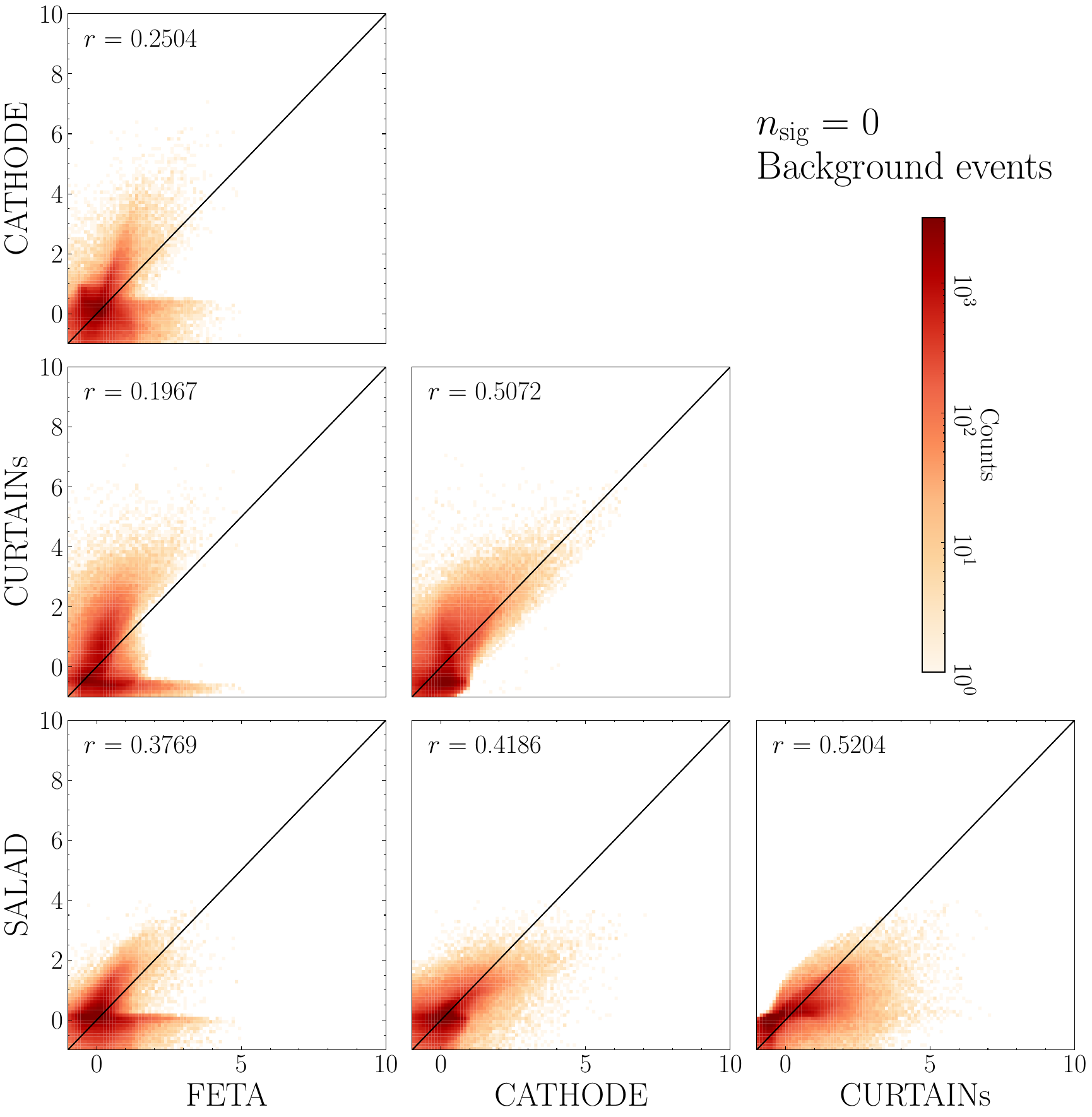}
    \caption{Scores for background events for a binary classifier trained to discriminate synthetic SM background from data with $n_\mathrm{sig} = 0$). Each axis represents a different method of SM sample generation. $r$ denotes the Pearson correlation coefficient, computed over all samples (not just the upper right quadrant). The scores are standardized so as to make it clear what events the binary classifier flags as the most anomalous. For each method, scores are averaged over 10 classifier runs.}
    \label{fig:corner_bkg_0_zoomed}
\end{figure}

Figure~\ref{fig:corner_bkg_0_zoomed} plots the classifier scores, averaged over 10 classifier runs, as evaluated on the test set's background events\footnote{See \App{app:more_plots} for the corresponding plots evaluated on the test set's signal events.}. Note that we plot the \textit{standardized} scores, since the binary classifier is trained to flag the most anomalous events with the highest scores. We also focus on the first quadrant of the coordinate plane, corresponding to the``anomaly regions" of the plots, or the highest regions in (standardized) score space where the classifier-flagged anomalies are expected to lie. In general, the classifier scores for the networks trained to discriminate detected data from each synthetic sample generation method do not appear to agree across methods: the scores are, for the most part, uncorrelated when evaluated on true background events. 

As a next task, we quantify the similarity across sample generation methods of events that are deemed ``signal-like" by the binary classifier. As our similarity metric, we consider the background events in the standard test set with classifier scores in the top $p$ percentile, i.e. the background events classified as the most ``signal-like", or most different to the synthetic background samples. Within the scope of an anomaly detection search, these top $p$-percentile events are exactly those that correspond to high-score, likely-to-be-anomalous events. 

\begin{figure}
    \centering
    \includegraphics[width=\linewidth]{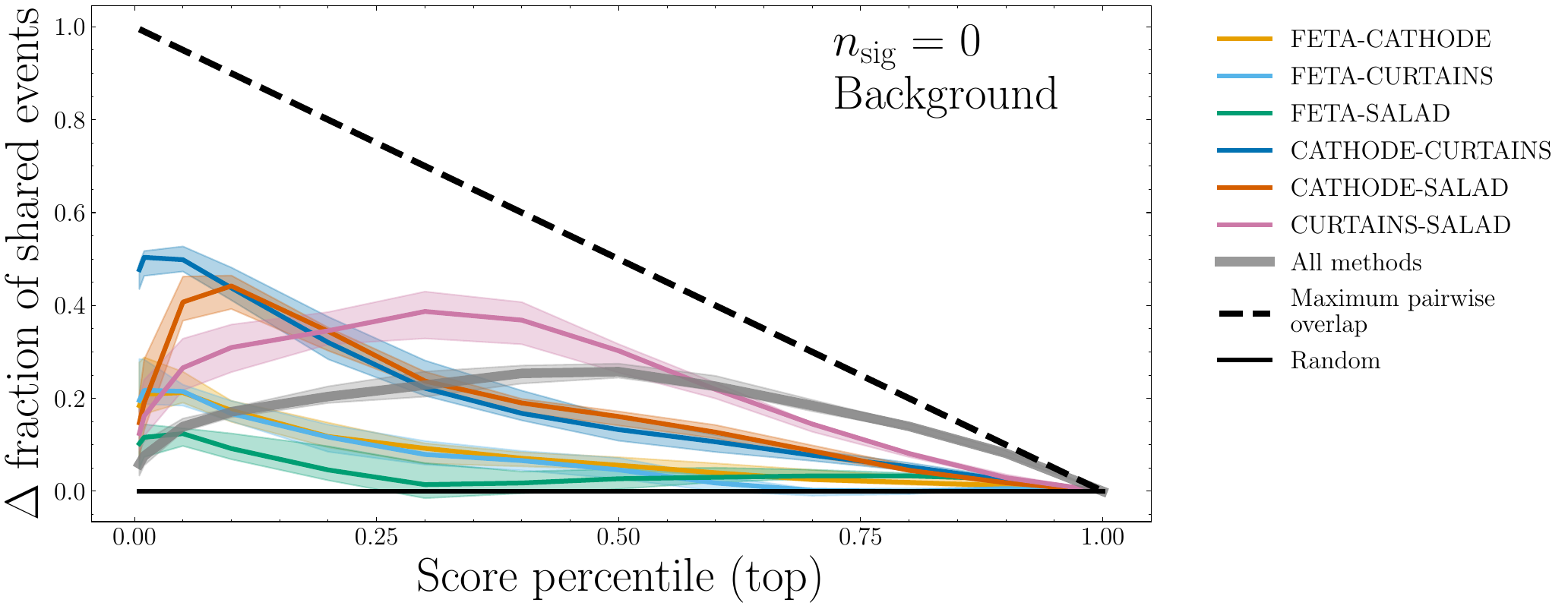}
    \caption{\label{fig:unification_bkg_0} Fractional overlap, with respect to a random-choice baseline, of the $p$th percentile of true background events classified as the most ``signal-like" between different methods of synthetic SM sample generation. Errorbands show a 68-percentile spread across the median and come from 100 repetitions of training 5-fold classifiers on the associated methods with different random seeds and ensembling scores over 10 repetitions. Note that the left-most point corresponds to a percentile of $p = 0.005$.}
\end{figure}

For each percentile $p$, we find the set of the top $p$ events independently for the classifiers trained on all four methods. We then calculate the overlaps between the sets of selected events between each pair of sample generation methods, then across all four methods. Note that for fully independent event selection methods, we would expect a fraction $p$ of shared events by chance; therefore for easier visualization, we plot the excess overlap with respect to this random baseline in \Fig{fig:unification_bkg_0} (subtracting the correct baseline for the overlap of all four methods). If the excess overlap is 0, then the performance overlap is no more than expected by random chance. 

For most combinations of synthetic sample generation methods, the amount of overlap between any two sample generation methods is slightly greater than what we would expect between two uncorrelated sets of random numbers, especially for larger score percentiles. However, there appears to be a large amount of similarity between \Cathode{} and \Salad{}, and between \Cathode{} and \Curtains{}, especially at small percentiles (all of which implies a somewhat smaller amount of similarity between \Feta{} and any other method). There is also large degree of similarity between \Curtains{} and \Salad{}, especially at larger percentiles. 

Another way to study the quality of the background-only samples relative to each other utilizes a multiclass classifier. This method of comparing different generative models was first introduced in~\citep{Lim:2022nft}  in the context of up-sampling hydrodynamical galaxy simulations. (See also \citep{Diefenbacher:2023vsw} for a subsequent application to comparing generative models for fast calorimeter simulation.) We use the same classifier architecture as before and only modify the output layer to yield 4 (softmaxed) numbers, which we interpret as the probabilities of the input samples belonging to one of the four methods\footnote{We checked a larger classifier architecture with more nodes and additional dropout, but the averaged results did not change with respect to the ones reported below.}. We also use a larger batch size of 1000, which we found necessary in order to get repeatable results. We use a subset of 400,000 samples from each of the four methods to train (60\%), test (20\%), and evaluate (20\%) the classifier. Samples from \Salad{} get their appropriate sample weight in training, testing, and evaluation. Since the average of these weights is about 0.98, we add an additional class weight of 1/0.98 = 1.021 to the \Salad{} samples to correct for the small imbalance. 

For final evaluation and comparison of the samples, we consider the average of the log posterior~\citep{Lim:2022nft}, which is defined as 
\begin{equation}
\label{eq:log-posterior}
LP(\text{model } i|\text{samples } j) = \frac{1}{N} \sum_{x_k \in j} \omega_k\log{p_{\text{model } i}(x_k)},
\end{equation}
evaluated on the held-out test datasets. Here, the sum includes all samples $x_k$ of the tested model $j\in$ (\Salad{}, \Cathode{}, \Curtains{}, \Feta{}), $\omega_k$ is the sample weight of the sample $x_k$, and $N$ is the number of samples in the set. Since individual runs tend to scatter, we average the log posteriors over 100 independent classifier trainings with different random seeds. A well-trained multiclass classifier should be able to identify the samples belonging to each model, therefore we would expect to have 

\begin{equation}
\label{eq:log-posterior.relation}
LP(\text{model } i|\text{samples } j=i) > LP(\text{model } i|\text{samples } j\neq i).
\end{equation}

\begin{figure}
    \centering
    \includegraphics[width = \textwidth]{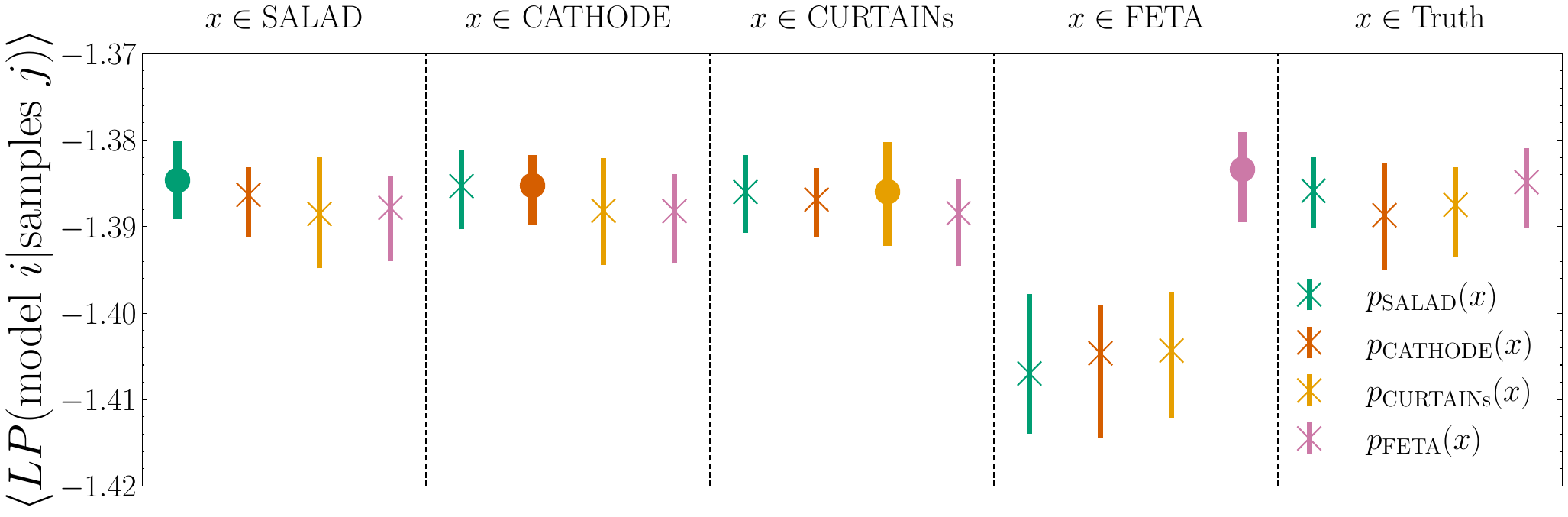}
    \caption{Average log posteriors $\langle LP(\text{model } i|\text{samples } j)\rangle $ of the multiclass classifier. The circle markers highlight the case $i=j$, and the cross markers indicate the cases $i\neq j$. ``Truth" designates the SR SM background events. Errorbars show a 68-percentile spread of the $LP$s of 100 independent retrainings (no score averaging is carried out).}
    \label{fig:logpost_methods}
\end{figure}

Indeed, this is what we observe in the first four columns of \Fig{fig:logpost_methods}. The probability of belonging to a given model is highest for samples that were generated with that model for \Salad{}, \Cathode{}, \Curtains{} (albeit only slightly for the latter two), and \Feta. These results are consistent with the previous similarity studies: \Salad{}, \Cathode{}, and \Curtains{} exhibit an above-average degree of similarity with each other, while \Feta{} appears to be more independent. To assess the question which of the methods produces artificial background closest to ``truth'' (the true SR background SM events), we evaluate the log posterior of Eq.~\eqref{eq:log-posterior} for samples from the truth dataset. We see in the right column of \Fig{fig:logpost_methods} that all four methods are essentially of equivalent quality, with their log posterior scores all well within each other's error bars. With respect to the truth, the ``\Feta{} anomaly" appears to be less pronounced. 
%these samples look on average most like \Salad{} and \Feta{} samples and least like \Cathode{} and \Curtains{}. However, the scores are very close to each other and the error bars overlap. 

\subsection{Adding in signal}

\begin{figure}
    \centering
    \includegraphics[width=.6\linewidth]{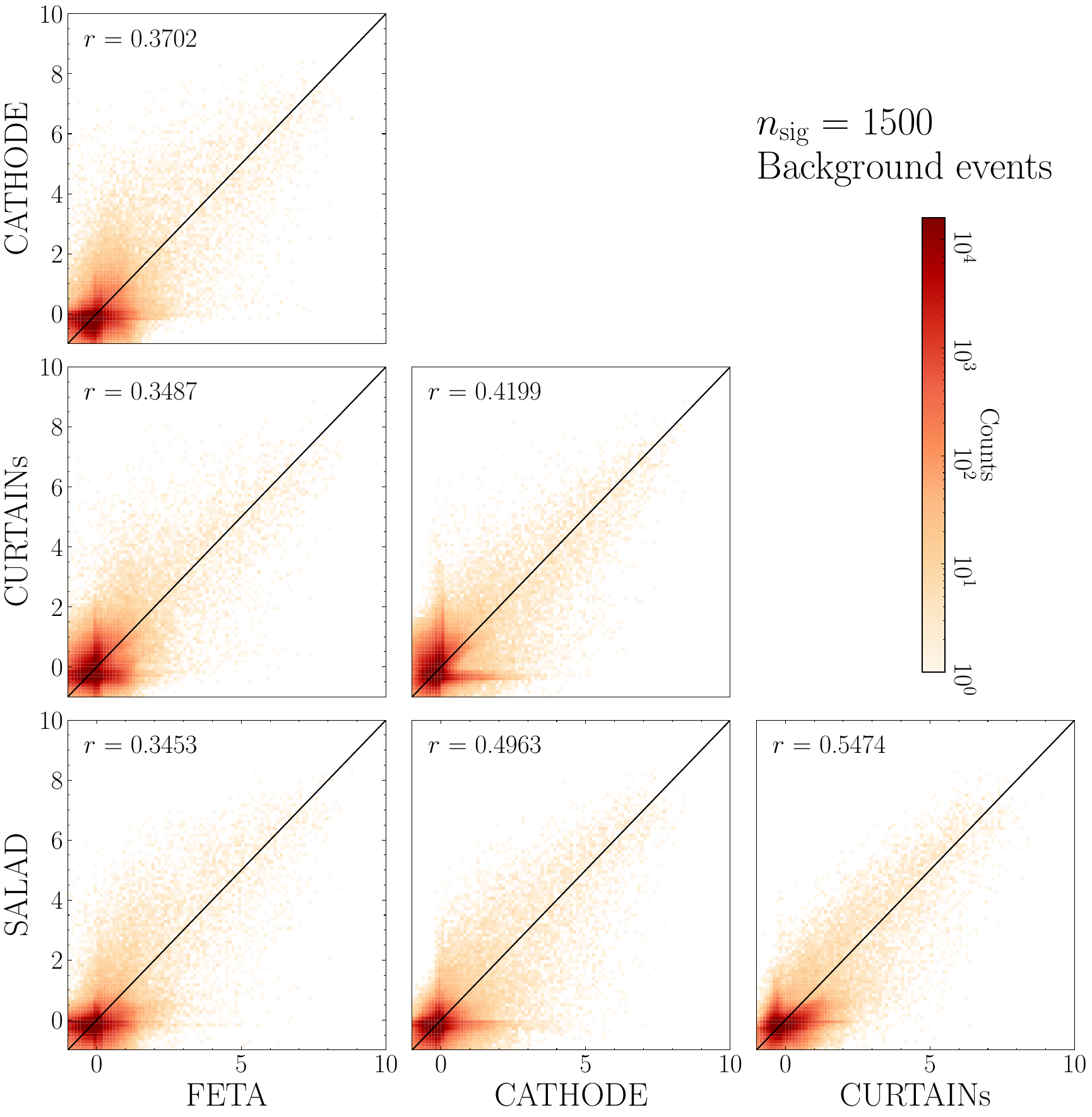}
    \caption{Scores for background events for a binary classifier trained to discriminate synthetic SM background from data with $n_\mathrm{sig} = 1500$, $S/\sqrt{B}$ = 3.24). Each axis represents a different method of SM sample generation. }
    \label{fig:corner_bkg_1500_zoomed}
\end{figure}

In \Figs{fig:corner_bkg_1500_zoomed}{fig:corner_sig_1500_zoomed}, we provide the scatterplots of the classifier scores evaluated on true background and true signal (respectively) events across different methods, this time for the case with injected signal: $n_\mathrm{sig} = 1500$ ($S/B = 0.93\%, S/\sqrt{B} = 3.24$). Note that we fully retrain all classifiers for this new signal injection.

\begin{figure}
    \centering
    \includegraphics[width=.6\linewidth]{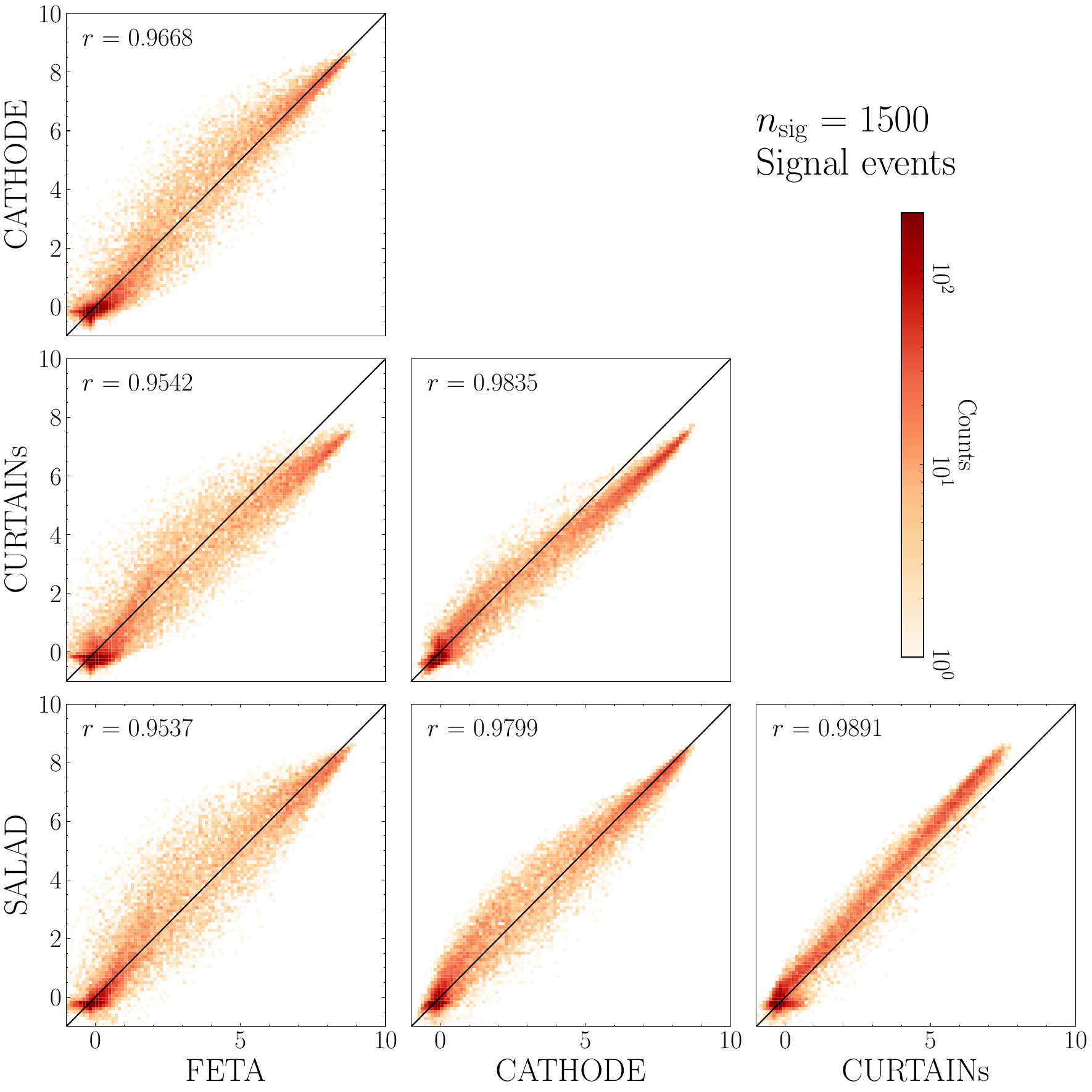}
    \caption{Scores for signal events for a binary classifier trained to discriminate synthetic SM background from data with $n_\mathrm{sig} = 1500$, $S/\sqrt{B}$ = 3.24). Each axis represents a different method of SM sample generation.}
    \label{fig:corner_sig_1500_zoomed}
\end{figure}

As shown in \Fig{fig:corner_bkg_1500_zoomed}, for this larger signal injection (as compared with 0 signal injection in Fig.~\ref{fig:corner_bkg_0_zoomed}), the correlation between classifier scores for background events across synthetic sample generation method is somewhat higher, especially for correlations involving \Salad{} or \Feta{}. In contrast, \Fig{fig:corner_sig_1500_zoomed} shows that the classifier scores for signal events are highly correlated between any two methods; all methods seem to agree on what anomalous events are. To summarize these results: we see that the classifier scores are rather uncorrelated on background events, but highly correlated on signal events. This might mean that the characteristics (i.e. the 5-dimensional non-mass feature space) of the synthetic background that is created differ non-trivially from method to method; there isn't overwhelming consensus on how to classify true background. However, all four of the methods produce background that is non-trivially different from true signal, at least different enough that classifiers can reliably distinguish background from signal.

\begin{figure}
  \begin{subfigure}[t]{.9\textwidth}
    \centering
    \includegraphics[width=\linewidth]{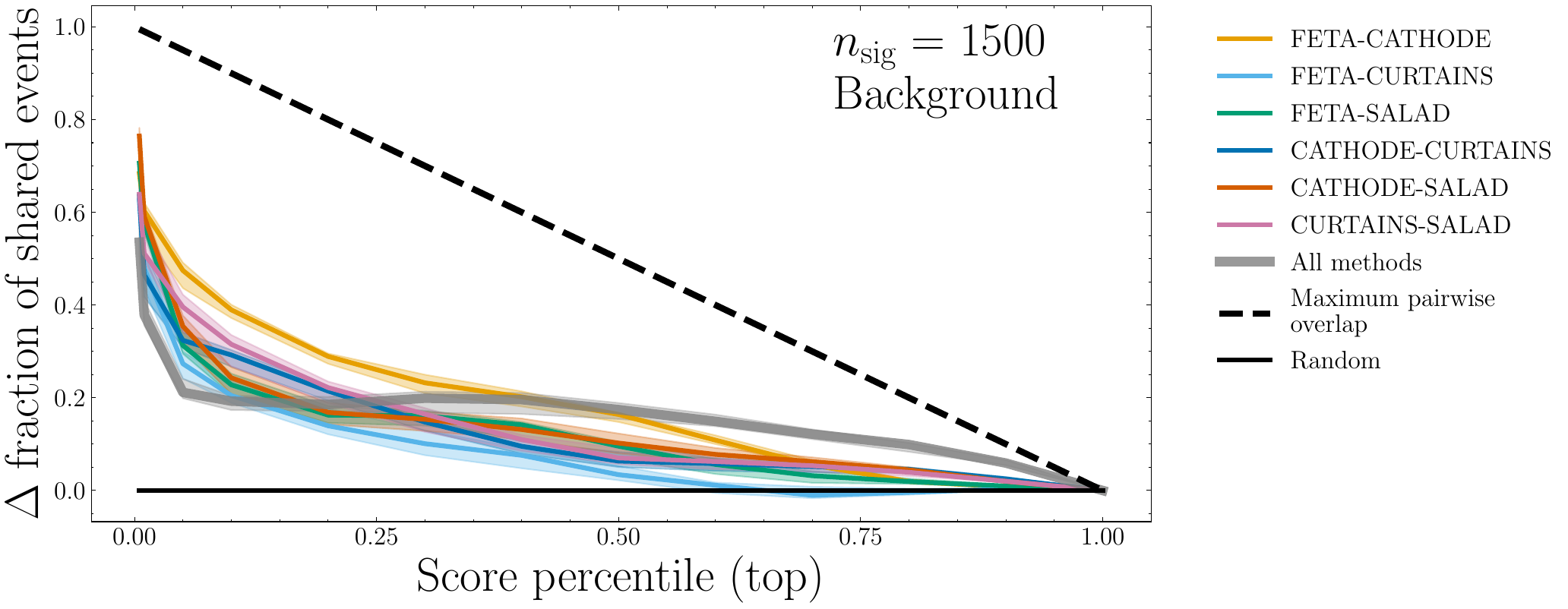}
    \caption{\label{fig:unification_bkg_1500}True background events.}
  \end{subfigure}
  \hfill
  \begin{subfigure}[t]{.9\textwidth}
    \centering
    \includegraphics[width=\linewidth]{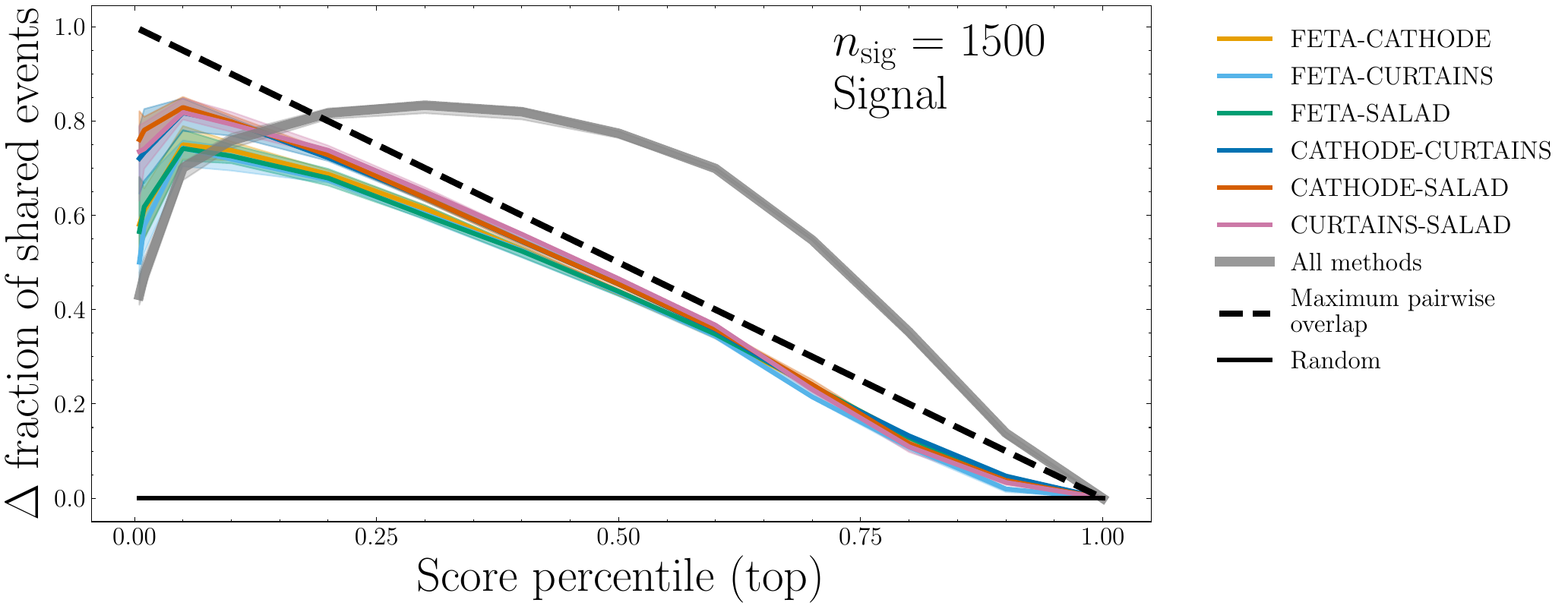}
    
    \caption{\label{fig:unification_sig_1500}True signal events. }
  \end{subfigure}
 
\caption{\label{fig:overlaps} Fractional overlap, with respect to a random-choice baseline, of the $p$th percentile of events classified as the most ``signal-like" between different methods of synthetic SM sample generation.}
\end{figure}

In \Fig{fig:overlaps}, we once again plot the overlaps of the top-$p$ percentile most ``signal-like" events for a training signal injection of $n = 1500$. For background events (\Fig{fig:unification_bkg_1500}), there is now a sizable amount of overlap between all pairs of methods at $p \lesssim 0.1$, though the overlap drops off quickly for larger percentiles. For the signal events (\Fig{fig:unification_sig_1500}), there is a significant excess of event overlaps between any two methods down to low-to-mid percentiles. This agrees with intuition: the natures of the synthetic SM samples may differ from method to method, but the hope is that they all differ significantly from a BSM resonance such that they can be used as a suitable background against which to discriminate the resonance. Importantly, there is an excess in event overlaps above random chance between all four methods across the board, at all percentiles.

\begin{figure}
  \begin{subfigure}[t]{\textwidth}
    \centering
    \includegraphics[width=.9\linewidth]{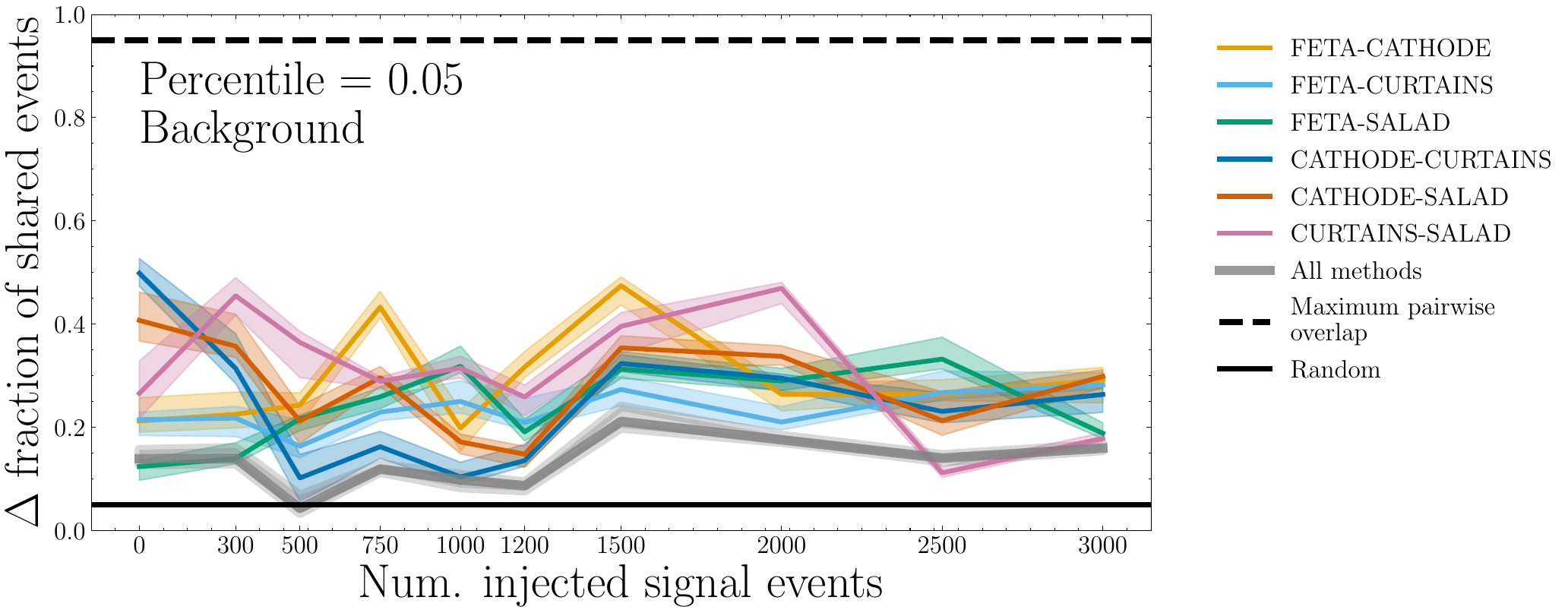}
    \caption{\label{fig:unification_percentile_5.0_bkg} True background events.}
  \end{subfigure}
  \hfill
  \begin{subfigure}[t]{\textwidth}
    \centering
    \includegraphics[width=.9\linewidth]{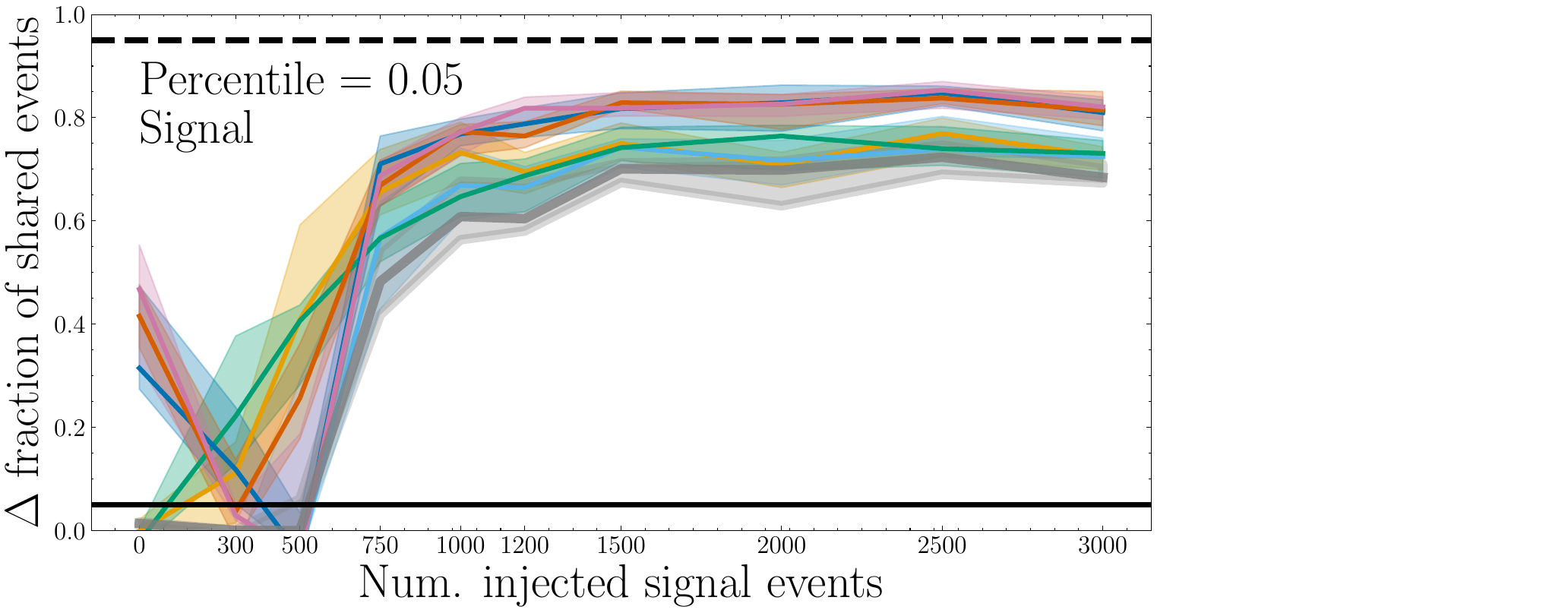}
    \caption{\label{fig:unification_percentile_5.0_sig}True signal events. For small signal injections ($n_\mathrm{sig} < 750)$, the negative values are likely a statistical fluctuation due to the small amount of signal present -- within errorbands, the values are consistent with 0.}
  \end{subfigure}
 
\caption{\label{fig:unification_percentile_5} Fractional overlap, with respect to a random-choice baseline, of the 5th percentile of events classified as the most ``signal-like" between different methods of synthetic SM sample generation, scanning over $n_\mathrm{sig}$.}
\end{figure}

In \Fig{fig:unification_percentile_5}, we consider a slightly different view of the percentile overlaps by fixing the percentile of the most ``signal-like" events and plotting this as a function of $n_\mathrm{sig}$ in the classifier training set. For the top 5 percentile of the most signal-like true background events, there appears to be slightly increasing similarity with the number of injected signal events $n_\mathrm{sig}$ across all four methods, but not between any two methods. For the top 5 percentile of the most signal-like true signal events, the agreement increases with $n_\mathrm{sig}$, leveling out at $n_\mathrm{sig} \approx 1200$. Put another way, the four methods considered here agree on what the 5\% most anomalous events are when trained to discriminate their own synthetic SM samples from a dataset containing signal injections as low as 0.62\%.

%\subsection{Discriminating between the samples}

%\rrm{Claudius study to be added?} moved up

\section{Combining the samples}
\label{sec:combining}

In this section, we investigate the extent to which \textit{combining} the synthetic samples can provide a more faithful approximation for SM background than taking samples from any of the individual generation methods alone. We have seen previously that classifiers trained to discriminate an individual method's synthetic samples from data tend to agree on what anomalous, signal-like events are more often than random. However, the agreement is not absolute. This might indicate that the events that each method's classifier are flagging as anomalous occupy slightly different parts of phase space. Therefore by combining the synthetic samples, we could hope to be more broadly sensitive to a larger phase space.

There are numerous ways to combine the sample generation methods, as the combination can in principle be done at one of many stages of an analysis. Additionally, one could imagine combining methods in a way that prioritizes one method over the other three. In this section, we will investigate the two most straightforward combinations that weight all four methods in the same way: first at the sample level, and second at the (classifier) score level.

To combine generation methods at the sample level, we take 250k samples from each method so that each contributes equally. We then train a binary classifier to discriminate the combined synthetic samples from the SR data, varying $n_\mathrm{sig}$ from 0 to 1500 (corresponding to $S/B$ in the SR of $0\%$ to $0.93\%$). We evaluate each classifier on the standard test set. To combine different methods at the score level, we simply average the score attributed to a given test set event over each of the four methods.

\begin{figure}
\centering
  \begin{subfigure}[t]{.49\textwidth}
    \centering
    \includegraphics[width=\linewidth]{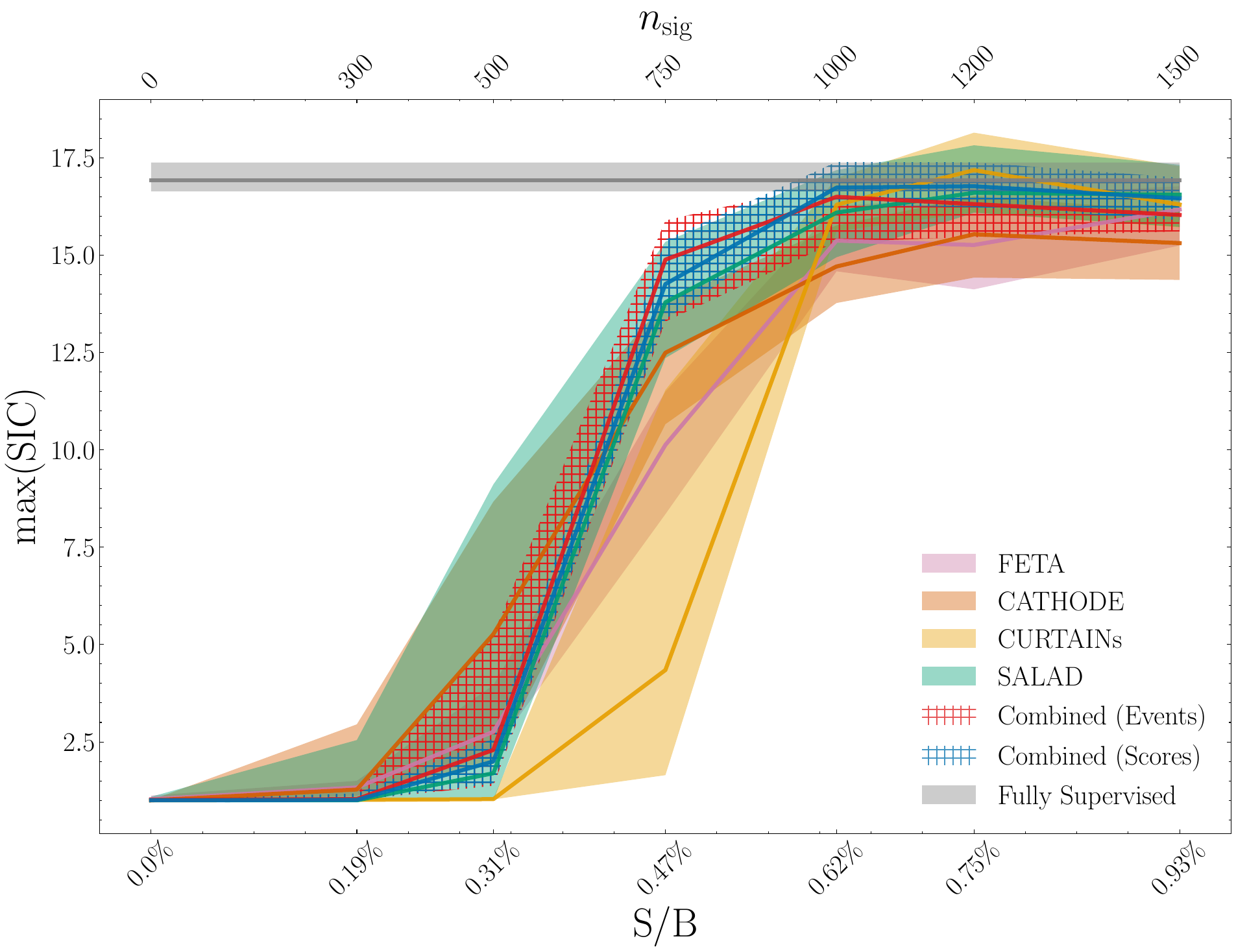}
    \caption{\label{fig:maxsic_allgen} Maximum of the Significance Improvement Characteristic.}
  \end{subfigure}
  \begin{subfigure}[t]{.49\textwidth}
    \centering
    \includegraphics[width=\linewidth]{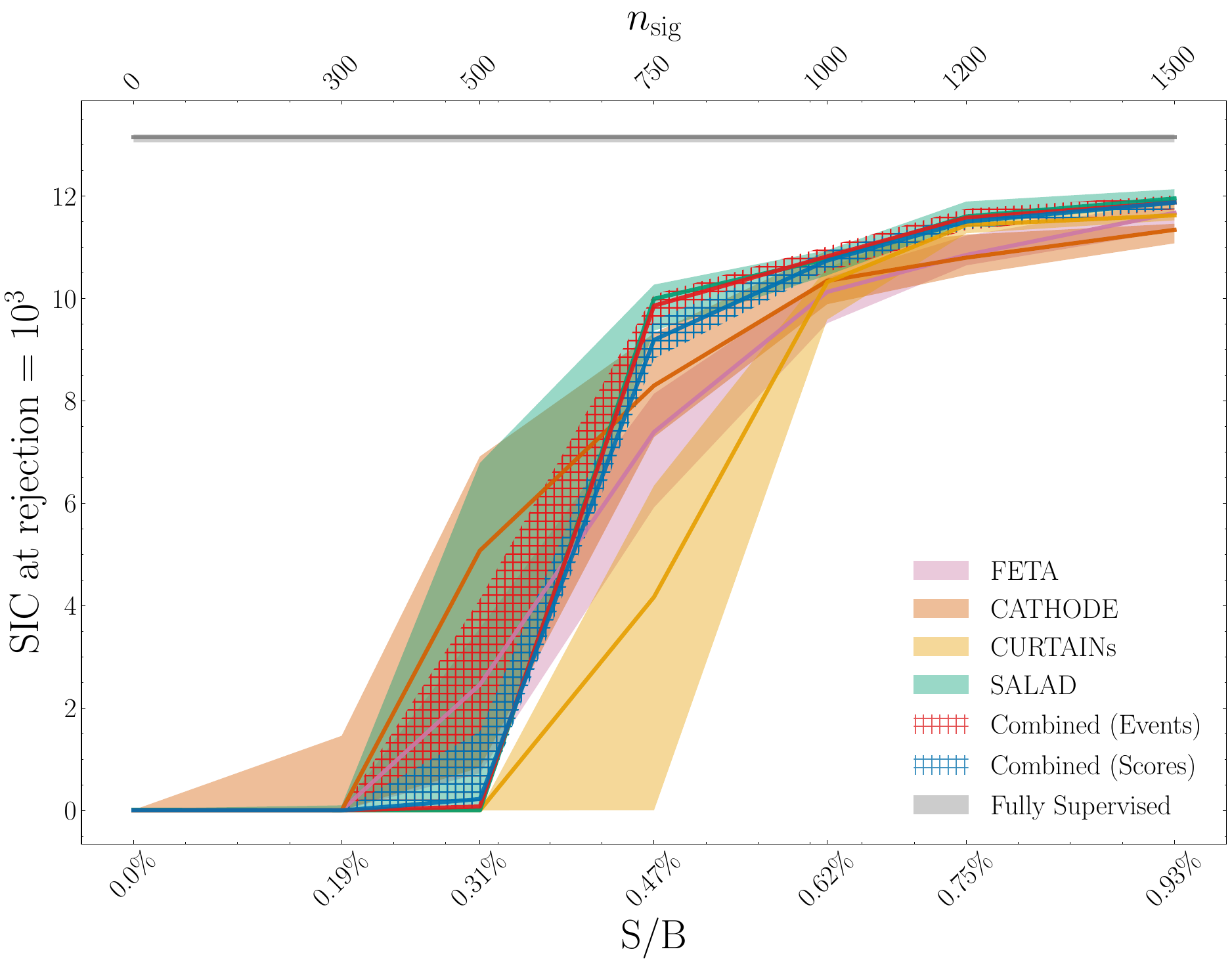}
   \caption{\label{fig:sic_at_rej_allgen} Significance Improvement Characteristic at a classifier rejection of $10^3$.}
  \end{subfigure}
\caption{\label{fig:bulk_metrics} Various metrics for a classifier trained to discriminate a combination of \Feta{}, \Cathode{}, and \Curtains{} synthetic SM samples from data over a range of $n_\mathrm{sig}$ values. Errorbands show a 68-percentile spread across the median and come from training a 5-fold classifier 100 times with different random seeds, over 3 independent generative model seeds, and ensembling scores over 10 runs.}
\end{figure}

To aggregate classifier runs, we train 100 such binary classifiers, average scores across ensembles of 10 runs, and generate classifier metric curves using the ensembled scores. This has the effect of tightening the errorbands for \Figs{fig:bulk_metrics}{fig:classifier_metrics_750}, making them easier to parse. For all methods, we apply a further level of aggregation by ensembling over the \textit{generator seed}. In other words, we create three instantiations of each generative ML model, repeat the analysis outlined in this and the previous paragraph, and amass all the classifier metric curves across the instantiations\footnote{This combination of generator seeds was found to give more robust results at the smallest signal injections. We explore this more in \App{app:robust}.}.

In \Fig{fig:bulk_metrics}, we provide summary plots across the range of tested $n_\mathrm{sig}$ values. In \Fig{fig:maxsic_allgen}, we calculate the classifier significance improvement characteristic (SIC) as a function of the signal efficiency, then take the maximum of the SIC. The max(SIC) gives the best multiplicative improvement to signal significance, corresponding to the best-motivated cut (which we do not know a priori). In \Fig{fig:sic_at_rej_allgen}, we plot the significance at a background rejection of $10^3$, which is less sensitive to the low-signal efficiency fluctuations of the max(SIC). Based on these metrics, the median performance of the combined synthetic samples is competitive with, but not necessarily better than, any of the individual sample generation methods; however, the spread of the combined samples is much tighter, implying greater stability.

\begin{figure}
\centering
  \begin{subfigure}[t]{.49\textwidth}
    \centering
    \includegraphics[width=\linewidth]{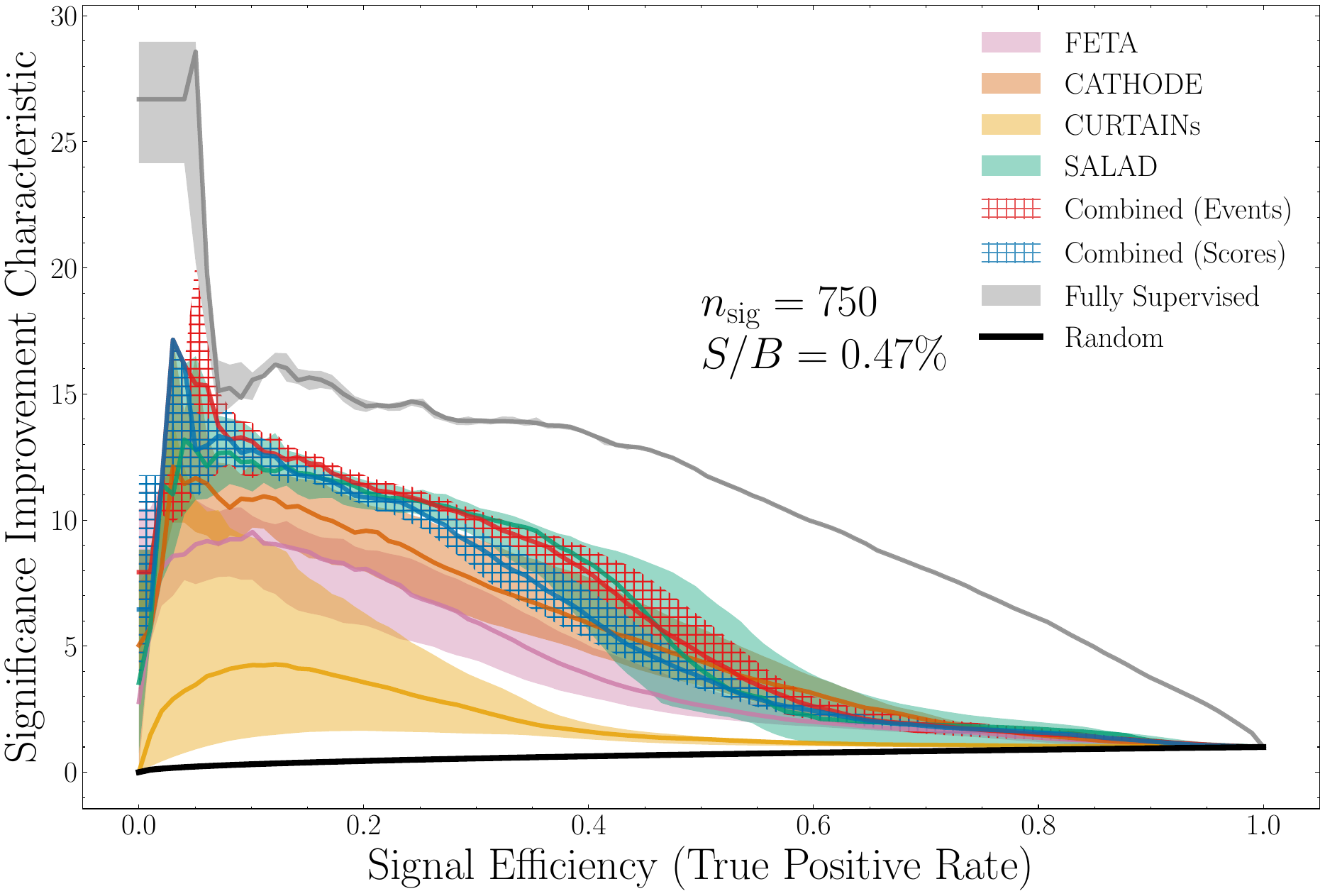}
    \caption{\label{fig:sic_allgen_750} Significance Improvement Characteristic plotted against the signal efficiency.}
  \end{subfigure}
  \begin{subfigure}[t]{.49\textwidth}
    \centering
    \includegraphics[width=\linewidth]{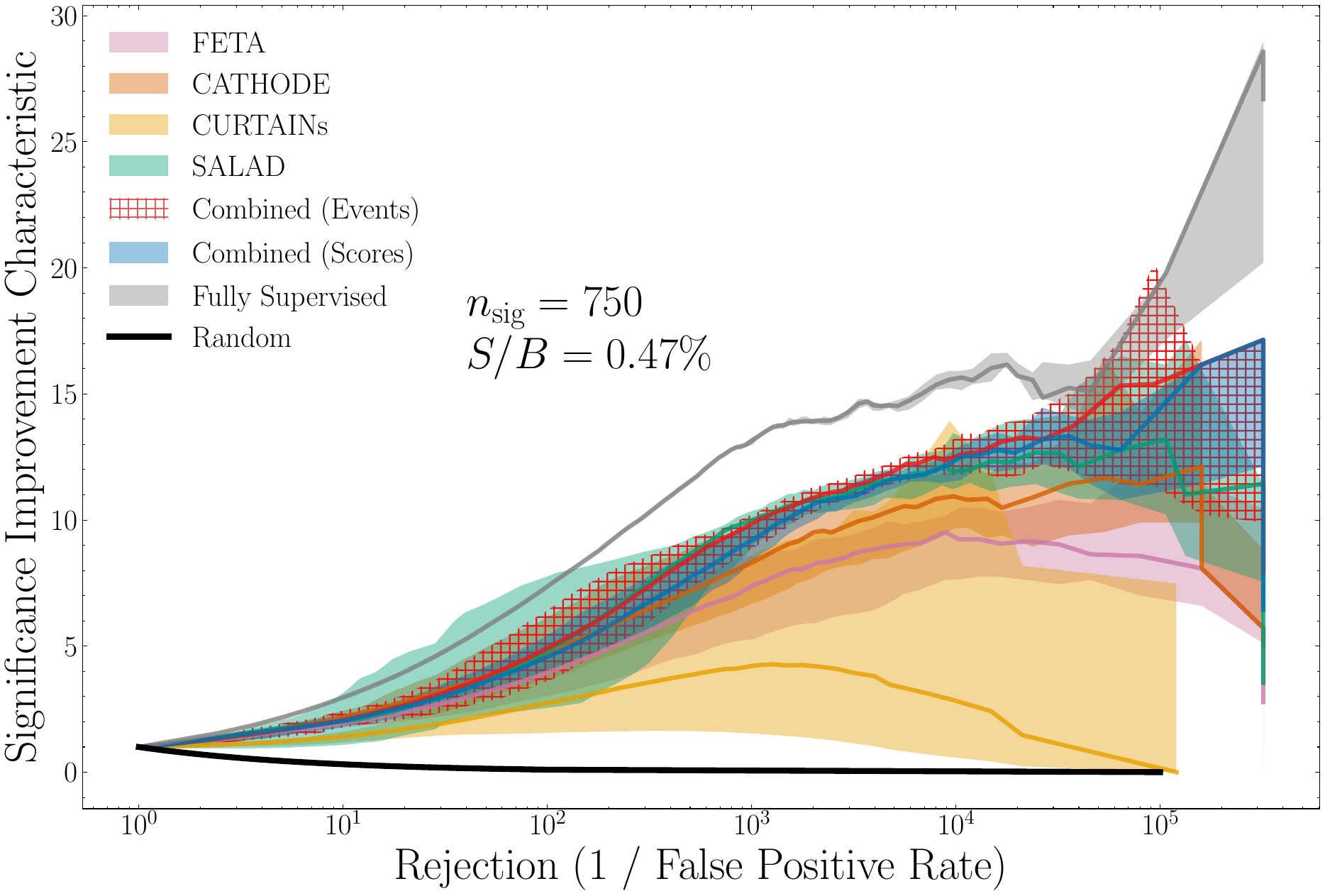}
   \caption{\label{fig:sic_vs_rej_allgen_750} Significance Improvement Characteristic plotted against classifier rejection.}
  \end{subfigure}
\caption{\label{fig:classifier_metrics_750} Various classifier metrics for a classifier trained to discriminate a combination of \Feta{}, \Cathode{}, and \Curtains{} synthetic SM samples from data with $n_\mathrm{sig} = 750$. Errorbands show a 68-percentile spread across the median and come from training a 5-fold classifier 100 times with different random seeds, over 3 independent generative model seeds, ensembling scores over 10 runs, and averaging classifier metrics over the ensembles.}
\end{figure}

The summary statistics alone may not be the most helpful gauge for performance in an AD task since we do not necessarily know the cut value corresponding to the max(SIC). In \Fig{fig:classifier_metrics_750}, we provide additional summary plots for the lowest signal injection $n_\mathrm{sig} = 750$ ($S/B = 0.47\%$) where using the combined samples leads to an improvement over using any individual method. While the two combination methods (i.e. at the sample and score levels) are comparable, the sample-level combination does appear to give better performance at most signal efficiencies\footnote{We provide equivalents to these plots computed at $n_\mathrm{sig} = 500$ in \Fig{fig:classifier_metrics_500} in \App{app:more_plots}, which shows that all methods fail to pick up on the signal.}.

Based on these plots, using the combined samples as the SM background leads to a classifier that is uniformly better (with respect to signal efficiency) at detecting the small amount of signal. This implies that when the score cutoff corresponding to the max(SIC) is unknown --- as it is in virtually all AD tasks --- \textit{combining} synthetic SM samples is the optimal strategy.

\section{Conclusions}
\label{sec:conclusions}

In this paper, we have explored four conceptually different methods of generating synthetic Standard Model (SM) background samples to be used for resonant anomaly detection (AD) tasks: \Salad{}, \Cathode{}, \Curtains{}, and \Feta{}. Each method uses a different means of generating a set of synthetic Standard Model samples to be used as a background set for resonant anomaly detection, but all use the same meta-format: shift in some way a sample of background-only events (pulled from an auxiliary dataset or background-only regions in data) to a signal region of interest, and search for a resonant anomaly within that region, such as by estimating an anomaly score based on the data-to-background likelihood ratio in non-$m$ features and cutting on it to enhance on the standard bump hunt procedure.

In general, the four construction methods produce synthetic SM samples that perform similarly when used for resonant AD tasks. Binary classifiers trained to discriminate SM samples from \Salad{}, \Cathode{}, \Curtains, and \Feta{} against data (SM background + injected signal) assign scores to the signal events that are generally correlated. Furthermore, the four methods agree on what the top $p$ percentile of the most ``signal-like" events are. While all methods perform similarly on their own with respect to being able to detect evidence of anomalous events as quantified by their max(SICs), \textit{combining} the four methods allows for a more sensitive AD tool at any given signal efficiency. This is especially useful in practice, when we do not know the optimal score cutoff corresponding to the max(SIC). We find that there is enough evidence to recommend that future AD tasks make use of this combined strategy for generating synthetic SM background samples.

With an eye towards future work: the LHC Olympics dataset is used as a benchmarking tool in the majority of resonant AD R\&D, but it represents just one resonant anomaly type out of a vast landscape. It is very possible that the results found in this report do not perfectly extend to other BSM particles, and therefore it would be useful to carry out similar tests of the SM generation methods on vastly different types of signal models. In that respect, it is interesting to remember the background (SM) -only studies in \Sec{sec:bkg_only}, which showed that the four methods considered in this work do seem to produce samples that cover non-overlapping regions of phase space. On a related vein, it would be worthwhile for future studies to explore how to make the anomaly-detecting CWoLa binary classifier signal-agnostic: it is standard in the field (especially for benchmarking studies) to optimize that classifier manually, which adds a degree of model-dependence into the anomaly detection procedure.

Finally, it would also be useful to consider other means of combining the synthetic samples, perhaps at the level of classifier metrics other than at the event-level or score-level, or in ways that prioritize one method in particular.

\section*{Acknowledgements}

BN and RM are supported by the U.S. Department of Energy (DOE), Office of Science under contract DE-AC02-05CH11231. RM is additionally supported by the National Science Foundation Graduate Research Fellowship Program under Grant No. DGE 2146752; any opinions, findings, and conclusions or recommendations expressed in this material are those of the authors and do not necessarily reflect the views of the National Science Foundation. The work of DSh was supported by DOE grant DOE-SC0010008. TG, JAR and DSe acknowledge funding through the SNSF Sinergia grant ``Robust Deep Density Models for High-Energy Particle Physics and Solar Flare Analysis (RODEM)'' with funding number CRSII5\_193716, and the SNSF project grant 200020\_212127 ``At the two upgrade frontiers: machine learning and the ITk Pixel detector''. GK and MS acknowledge support by the Deutsche Forschungsgemeinschaft under Germany’s Excellence Strategy – EXC 2121  Quantum Universe – 390833306.\ 
The work of MS was supported by BMBF grant 05H21GUCC1. CK would like to thank the Baden-W\"urttemberg-Stiftung for financing through the program \textsl{Internationale Spitzenforschung}, pro\-ject \textsl{Uncertainties – Teaching AI its Limits} (BWST\_IF2020-010).

\section*{Code avalability}

%The processed datasets used to generate all plots in this paper are provided at \citep{dataset}. 
All analysis code can be found at \url{https://github.com/rmastand/synthetic_SM_AD/tree/main}.

\bibliography{main.bib}

\appendix

\section{Robustness of classifier scores against network initialization}
\label{app:robust}

One might hope that the nature of the synthetic samples produced by each of the four generation methods considered in this report is robust in the sense that the samples produce the same results despite variances in initialization of neural networks. To this end, there are two sources of randomness: initialization of the binary classifier architectures, and initialization of the generator-level architectures, i.e. the networks used to generate the synthetic samples.

\subsection{Binary classifier initialization}

\begin{figure}
\centering
  \begin{subfigure}[t]{.6\textwidth}
    \centering
    \includegraphics[width=\linewidth]{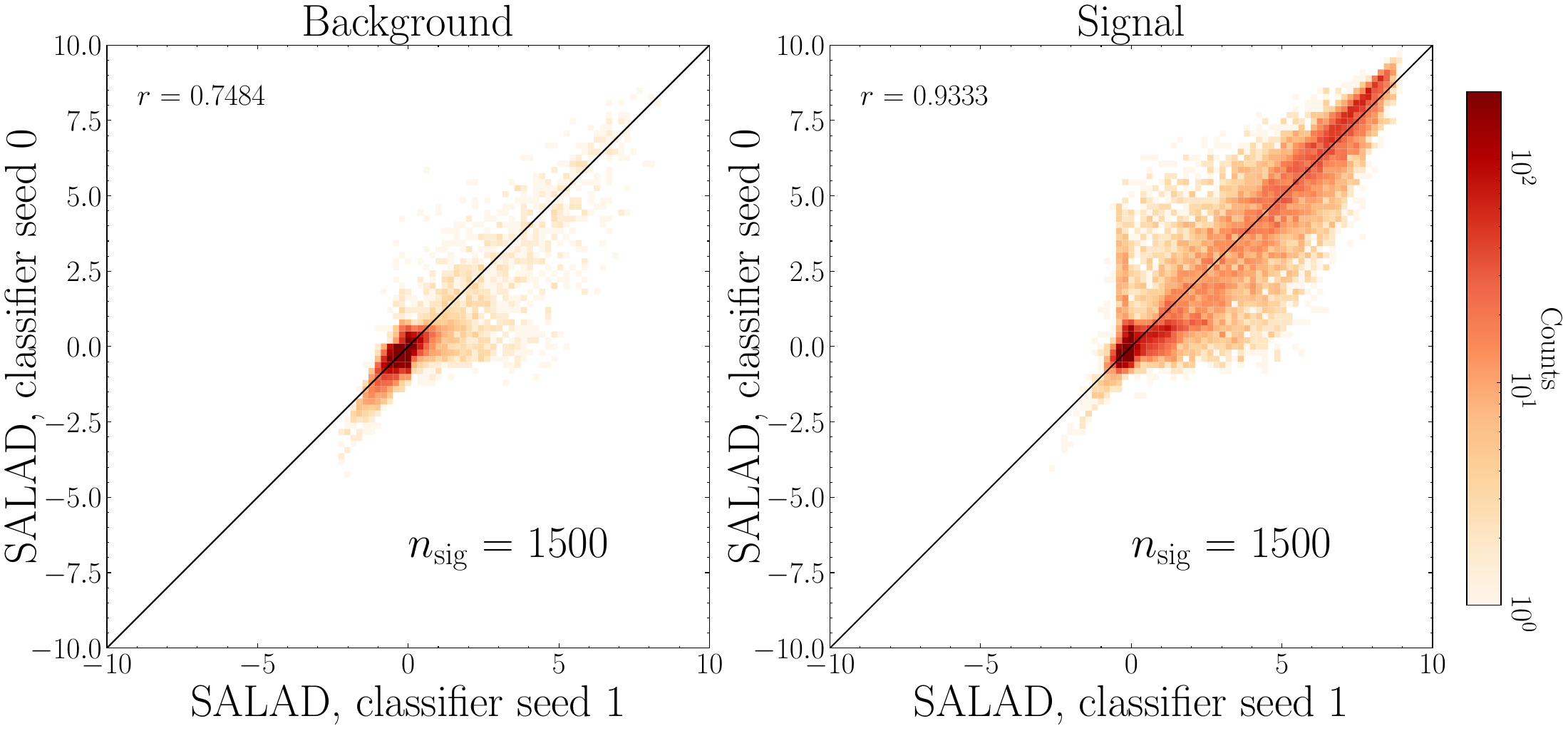}
    \caption{\Salad{} against \Salad{}.}
  \end{subfigure}
    \hfill
  \begin{subfigure}[t]{.6\textwidth}
    \centering
    \includegraphics[width=\linewidth]{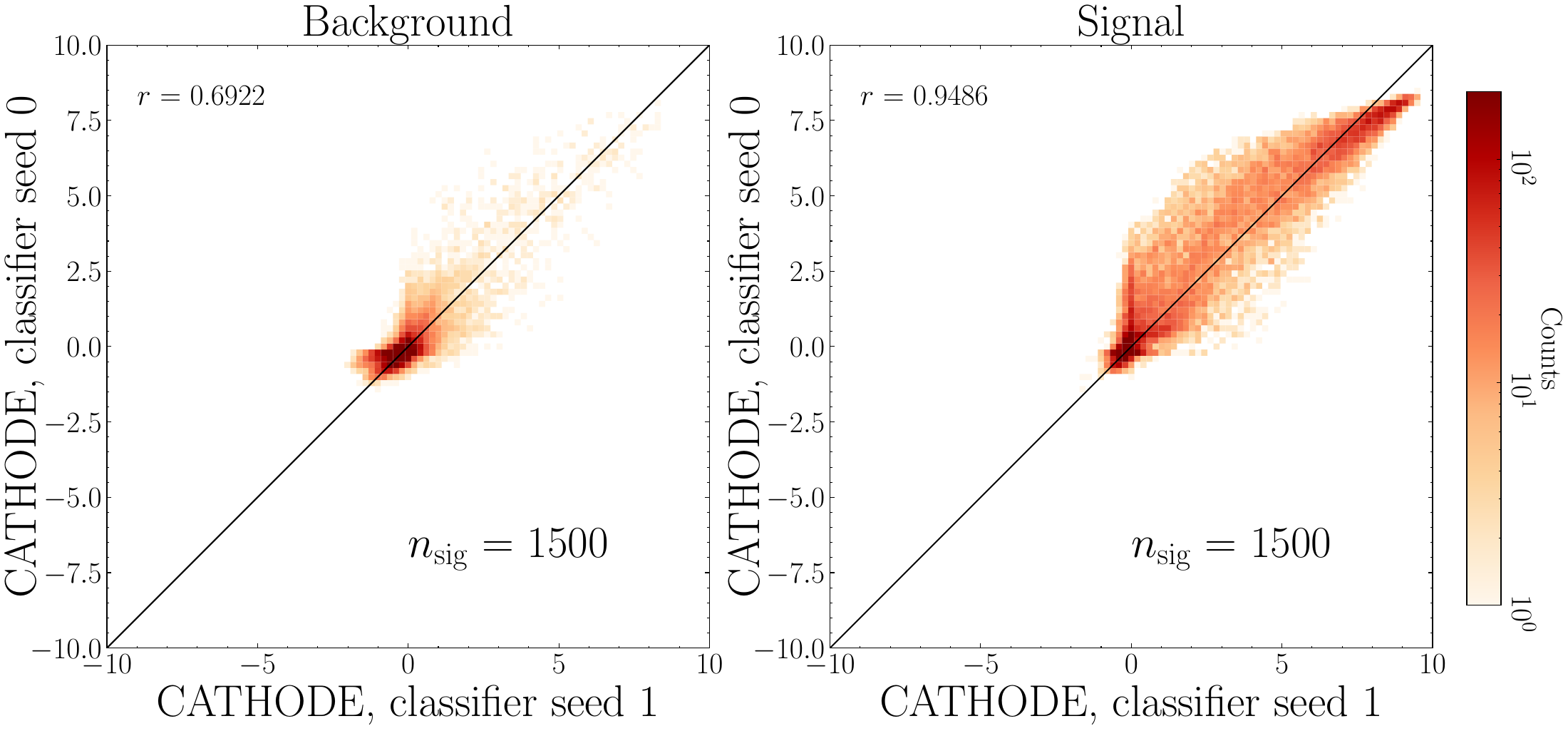}
   \caption{\Cathode{} against \Cathode{}.}
  \end{subfigure}
\hfill
  \begin{subfigure}[t]{.6\textwidth}
    \centering
    \includegraphics[width=\linewidth]{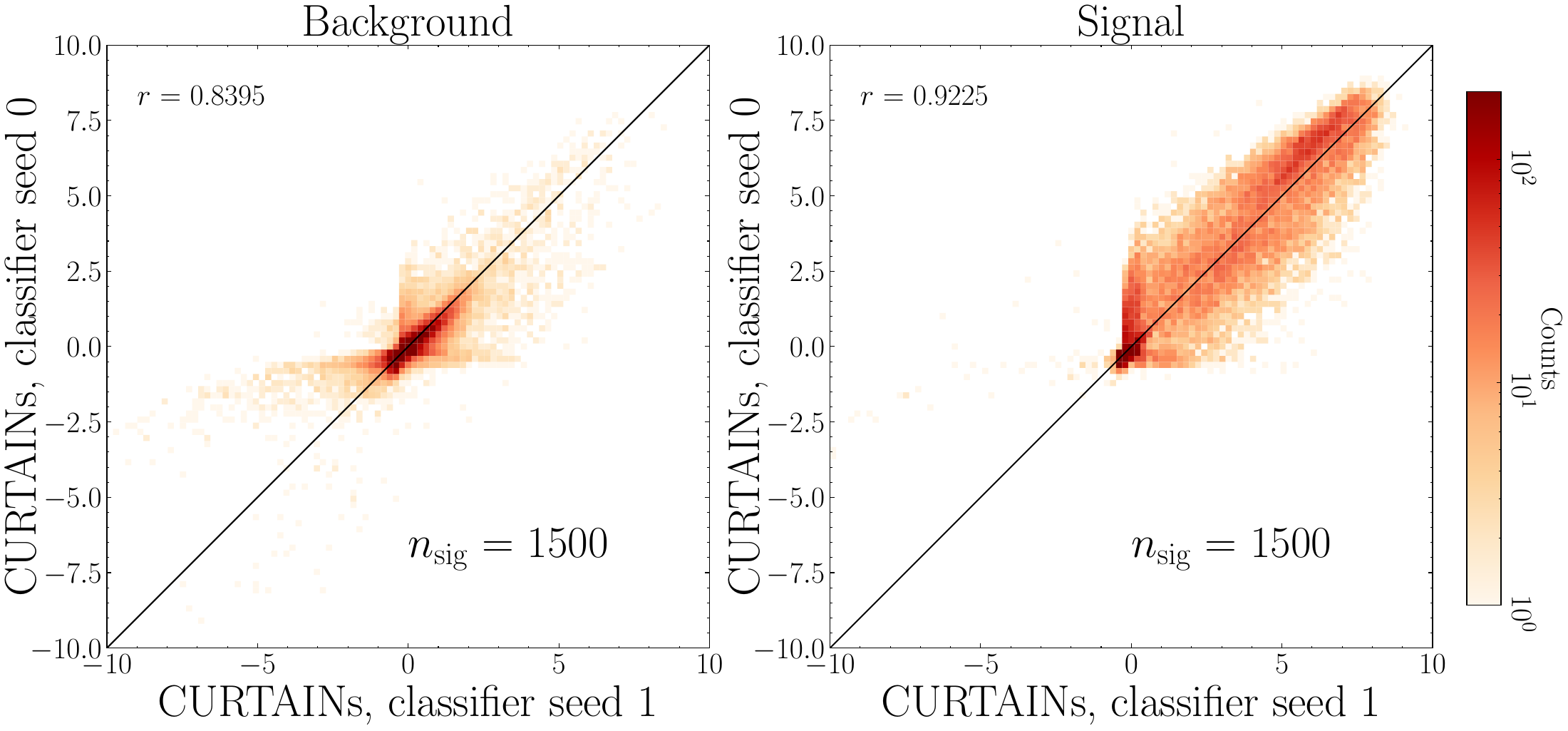}
   \caption{\Curtains{} against \Curtains{}.}
  \end{subfigure}
  \hfill
  \begin{subfigure}[t]{.6\textwidth}
    \centering
    \includegraphics[width=\linewidth]{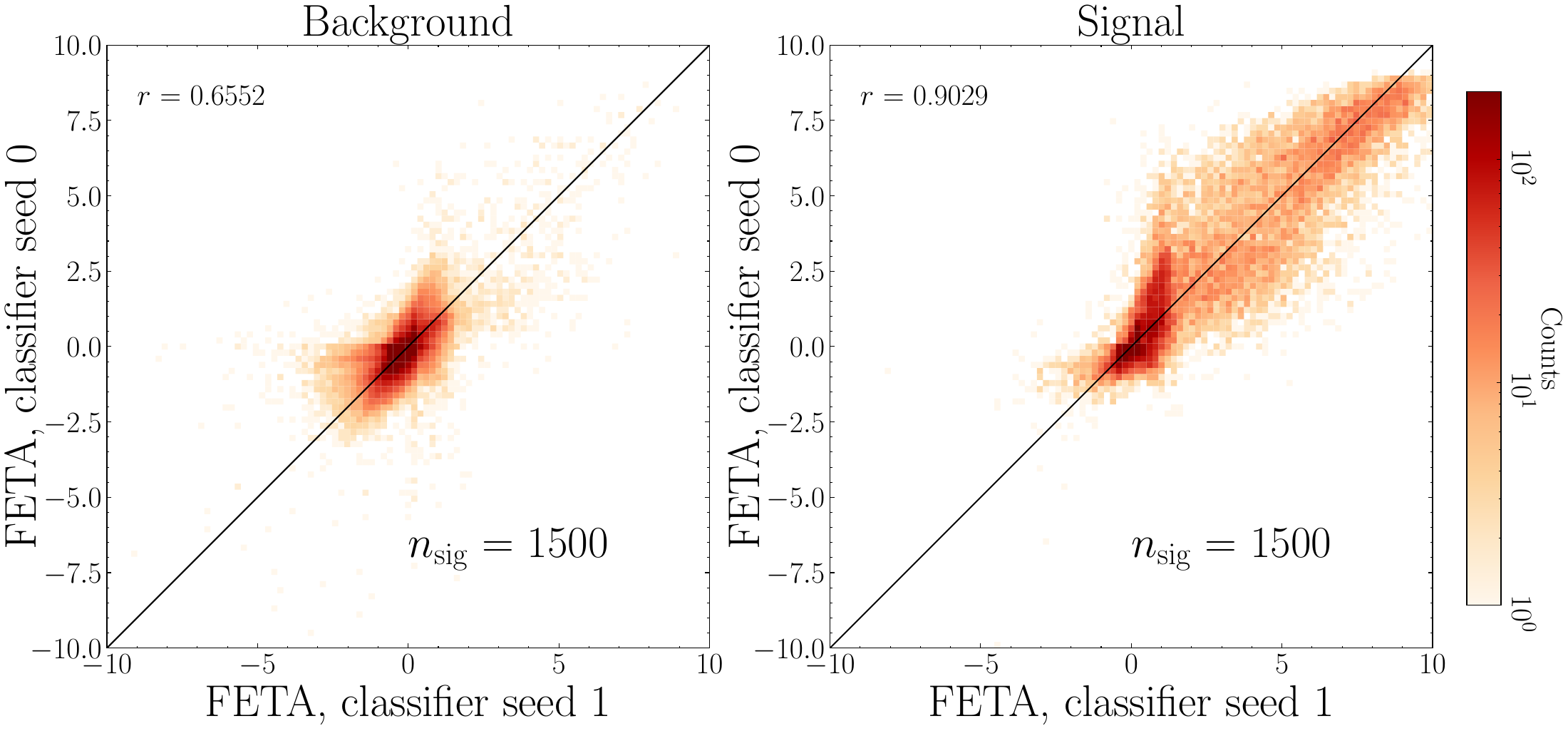}
   \caption{\Feta{} against \Feta{}.}
  \end{subfigure}
\caption{ Classifier scores for a binary classifier trained to discriminate synthetic SM background from  data with $n_\mathrm{sig} = 1500$). Each axis represents a different binary classifier random seed for the same sample generation method.}
\label{fig:seed_v_seed}
\end{figure}

We first gauge the robustness of scores against the binary classifier seed. In \Fig{fig:seed_v_seed}, we plot the classifier scores derived from samples of the \textit{same} method, but for classifiers trained starting with a different random seed, on a dataset with $n_\mathrm{sig} = 1500$ injected signal events. 

These plots illustrate that the scores output by the random classifier are relatively robust with respect to the training procedure when trained on a single sample generation method: the correlation between different random seeds is greater than 90\% for all methods for $n_\mathrm{sig} = 1500$. For smaller signal injections, the score correlation is reduced (especially below $n_\mathrm{sig} = 300$), but does remain positive. 

In the main text of this paper, we elect to show results derived from scores averaged over 10 classifier trials (i.e. with different random seeds). This choice helps to stabilize the results, such that we are not comparing an uncharacteristically good classifier run for one method with an uncharacteristically bad classifier run for another.

\subsection{Generative network initialization}

\begin{figure}
\centering
  \begin{subfigure}[t]{.6\textwidth}
    \centering
    \includegraphics[width=\linewidth]{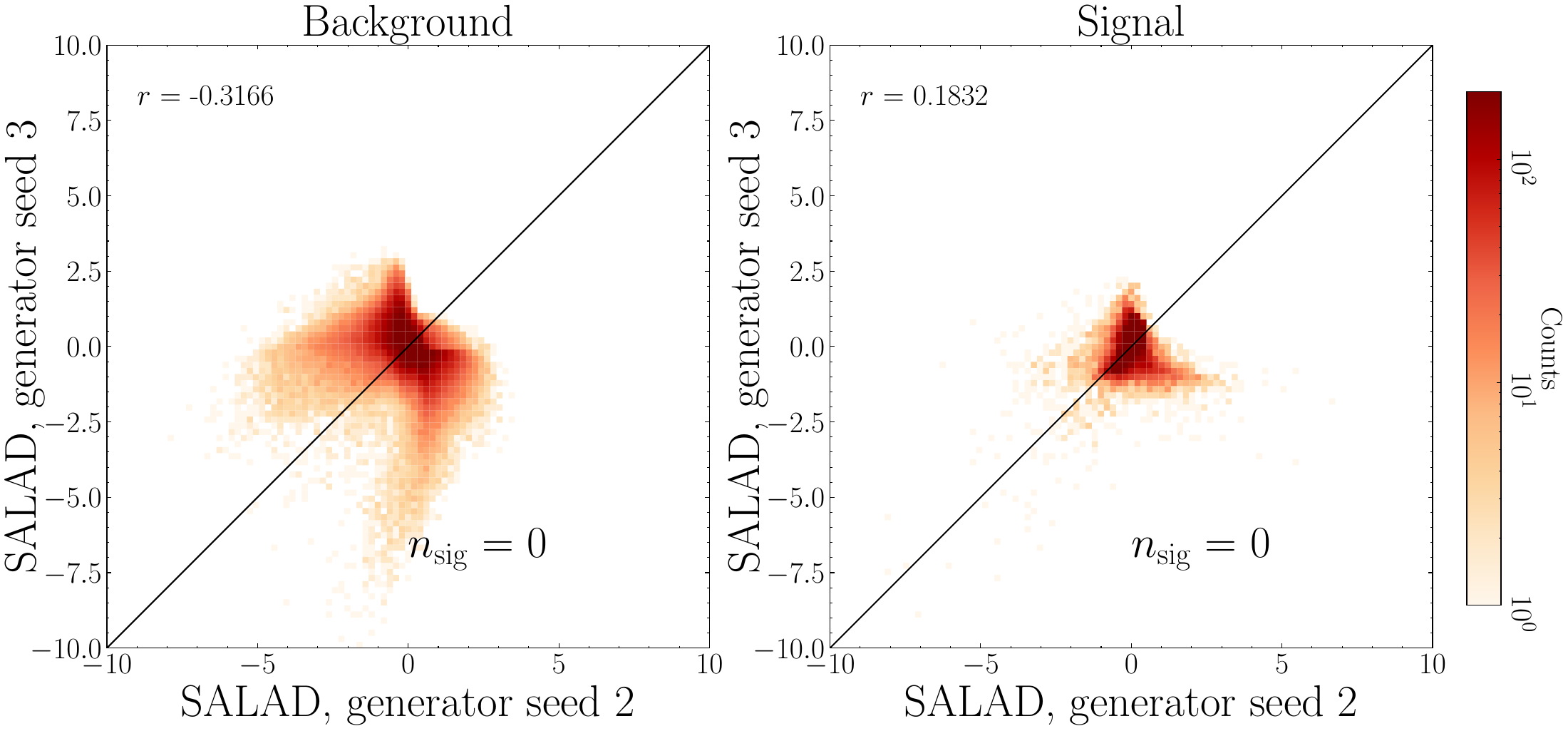}
    \caption{\Salad{} against \Salad{}.}
  \end{subfigure}
    \hfill
  \begin{subfigure}[t]{.6\textwidth}
    \centering
    \includegraphics[width=\linewidth]{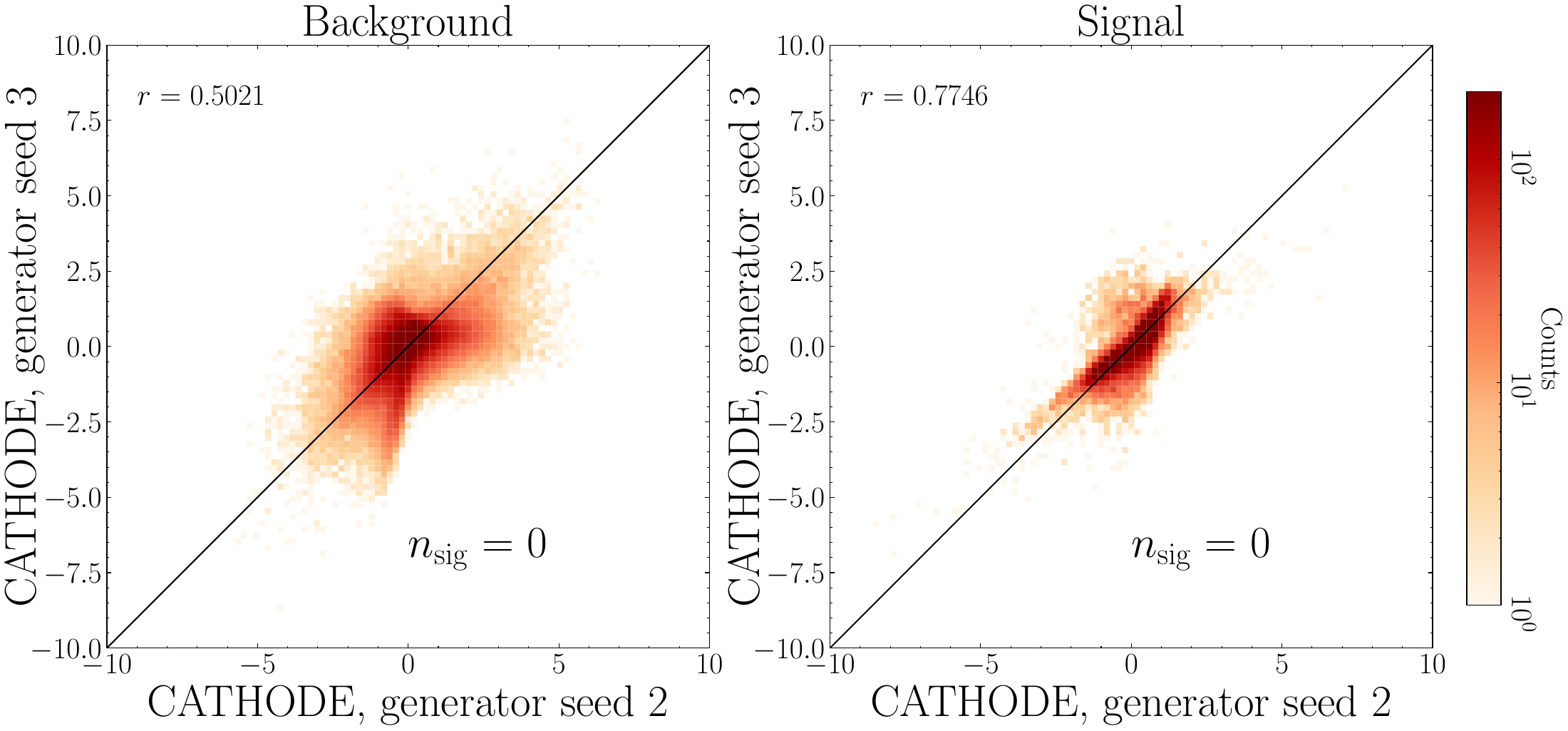}
   \caption{\Cathode{} against \Cathode{}.}
  \end{subfigure}
\hfill
  \begin{subfigure}[t]{.6\textwidth}
    \centering
    \includegraphics[width=\linewidth]{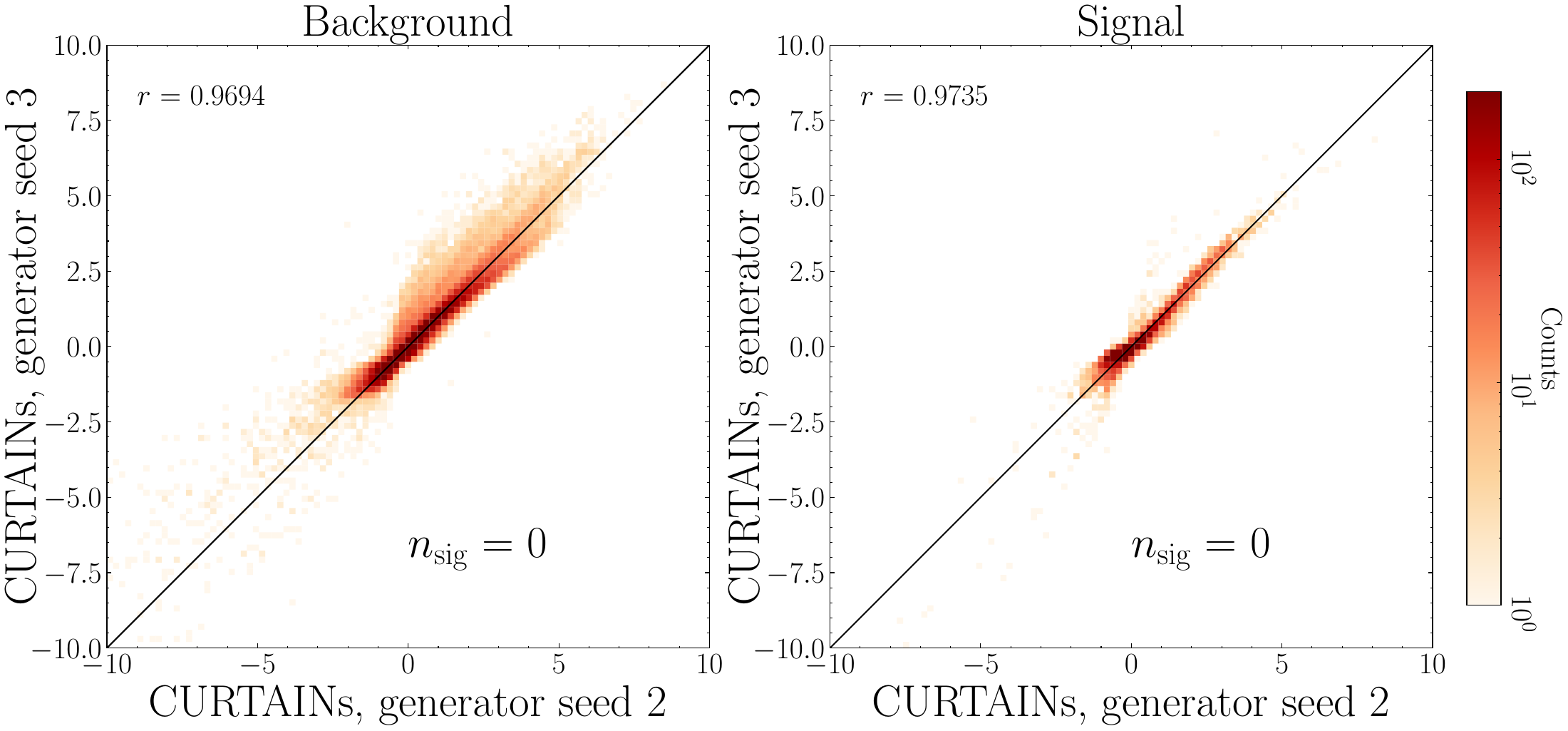}
   \caption{\Curtains{} against \Curtains{}.}
  \end{subfigure}
  \hfill
  \begin{subfigure}[t]{.6\textwidth}
    \centering
    \includegraphics[width=\linewidth]{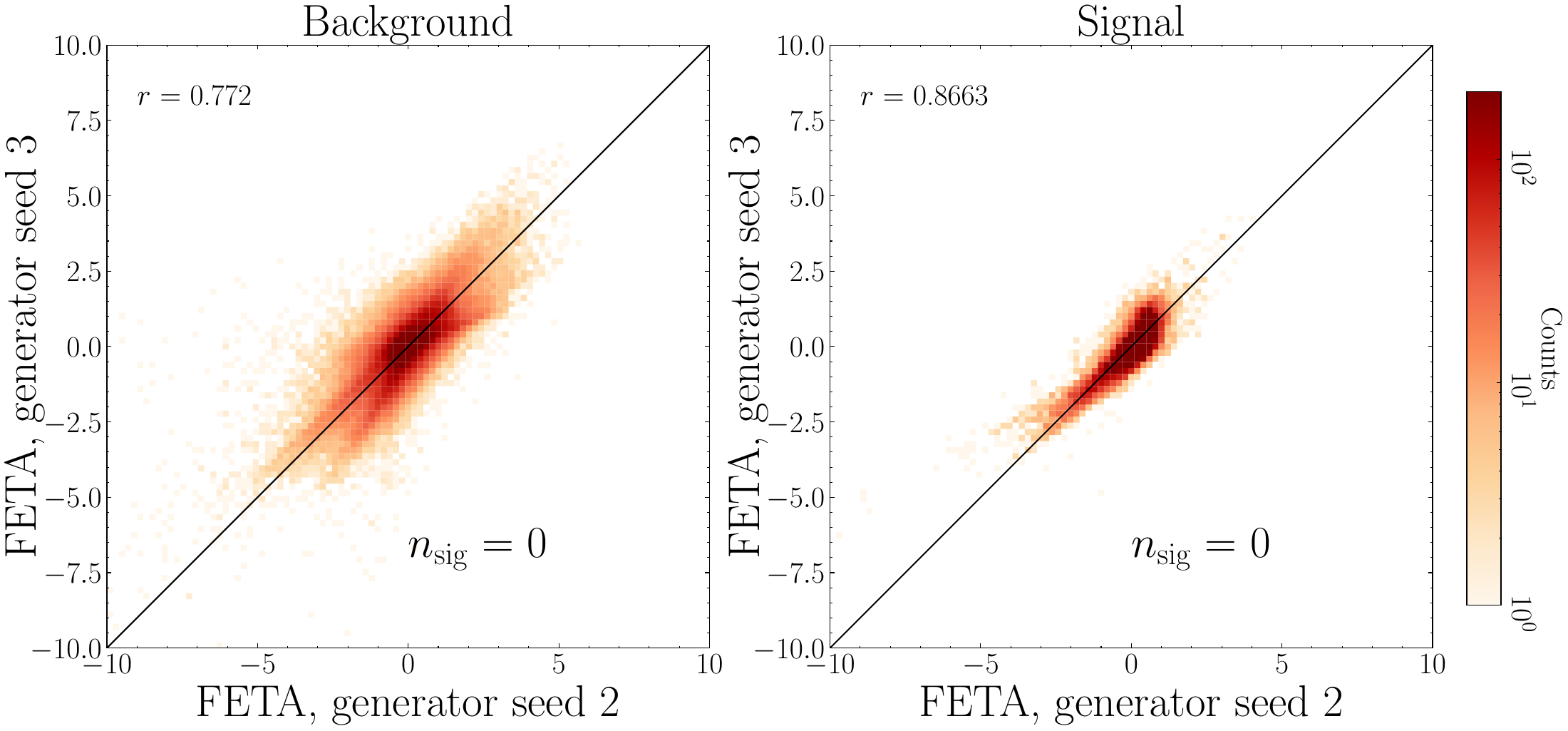}
   \caption{\Feta{} against \Feta{}.}
  \end{subfigure}
\caption{ Classifier scores for a binary classifier trained to discriminate synthetic SM background from  data with $n_\mathrm{sig} = 0$). Each axis represents a different generator architecture random seed for the given sample generation method. Scores are averaged over 10 different binary classifier instantiations.}
\label{fig:gen_v_gen}
\end{figure}

The question of score correlation across generative model seeds is an interesting one to ask particularly in the zero-signal injection case, when the binary classifiers can not leverage the characteristic observable distributions of signal events to make up for the differences between the different types of background that each synthetic sample generation method creates. In this case, it is useful to know if each sample generation method consistently produces similarly-functioning synthetic background samples. In \Fig{fig:gen_v_gen}, we plot the classifier scores derived from samples of the \textit{same} method, but for generative models initialized starting with a different random seed, on a dataset with $n_\mathrm{sig} = 0$ injected signal events. Note that we average over 10 instantiations of binary classifiers initialized with different seeds. 

We see that for no injected signal, there is a good deal of correlation for the classifier scores of the true signal events between the scores derived from differently seeded generator models. However, the correlation for the scores of the true background events is only mild for \Cathode{} samples, and is negative for \Salad{} samples (note that the correlations jump to 74\% and 72\%, respectively, for the case where $n_\mathrm{sig} = 1500$). Therefore a maximally robust study should make use of averaging over generator initializations in addition to over binary classifier initializations, or perhaps by modifying the synthetic sample generator training procedures to reduce sensitivity to network instantiation.

\subsection{Using non-robustness to indicate a breakdown of \textsc{CWoLa}}
\label{sec:cwola_breakdown}

\begin{figure}
\centering
  \begin{subfigure}[t]{.49\textwidth}
    \centering
    \includegraphics[width=\linewidth]{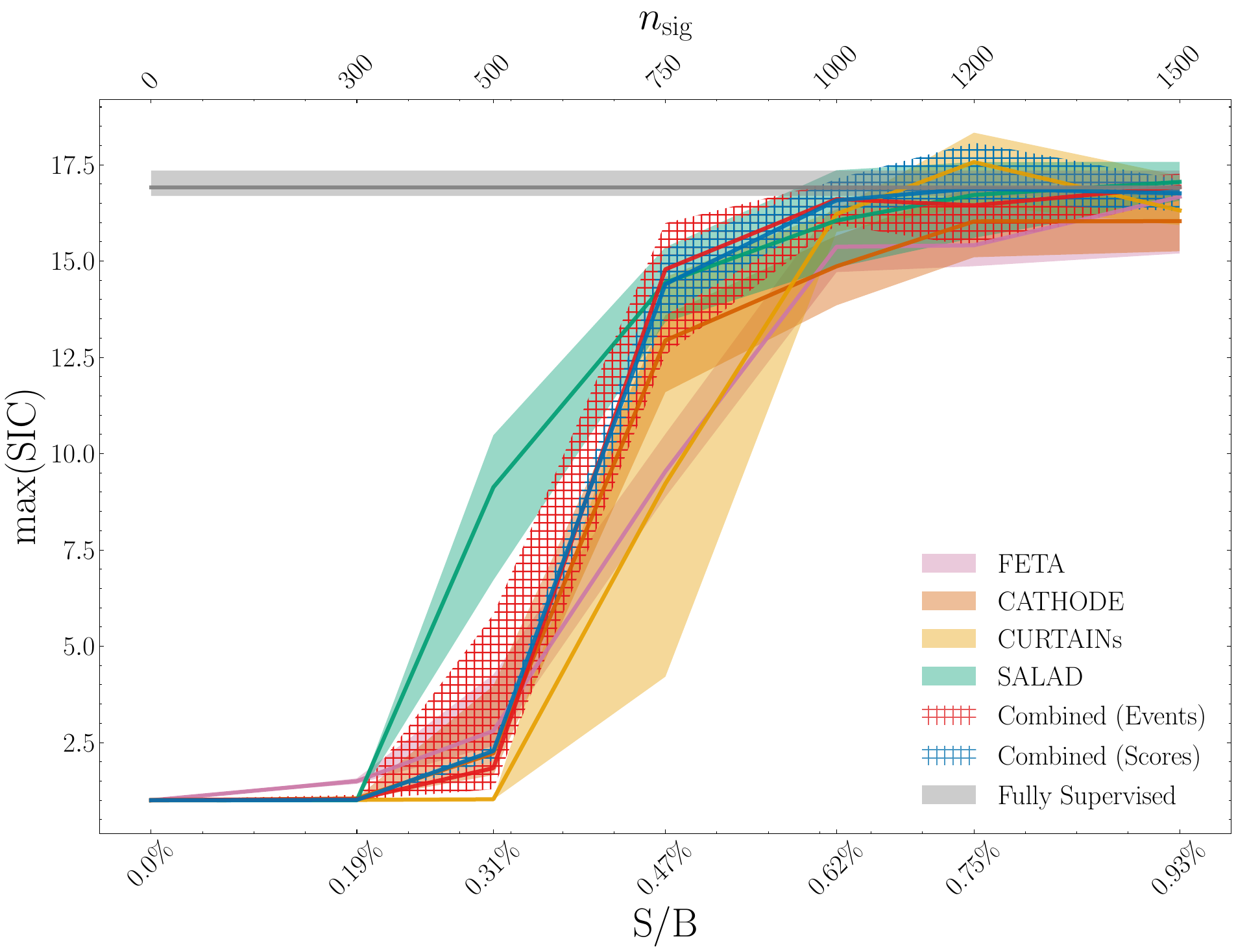}
    \caption{\label{fig:maxsic_genseed2} Maximum of the SIC, all generator seeds set to ``2".}
  \end{subfigure}
  \begin{subfigure}[t]{.49\textwidth}
    \centering
    \includegraphics[width=\linewidth]{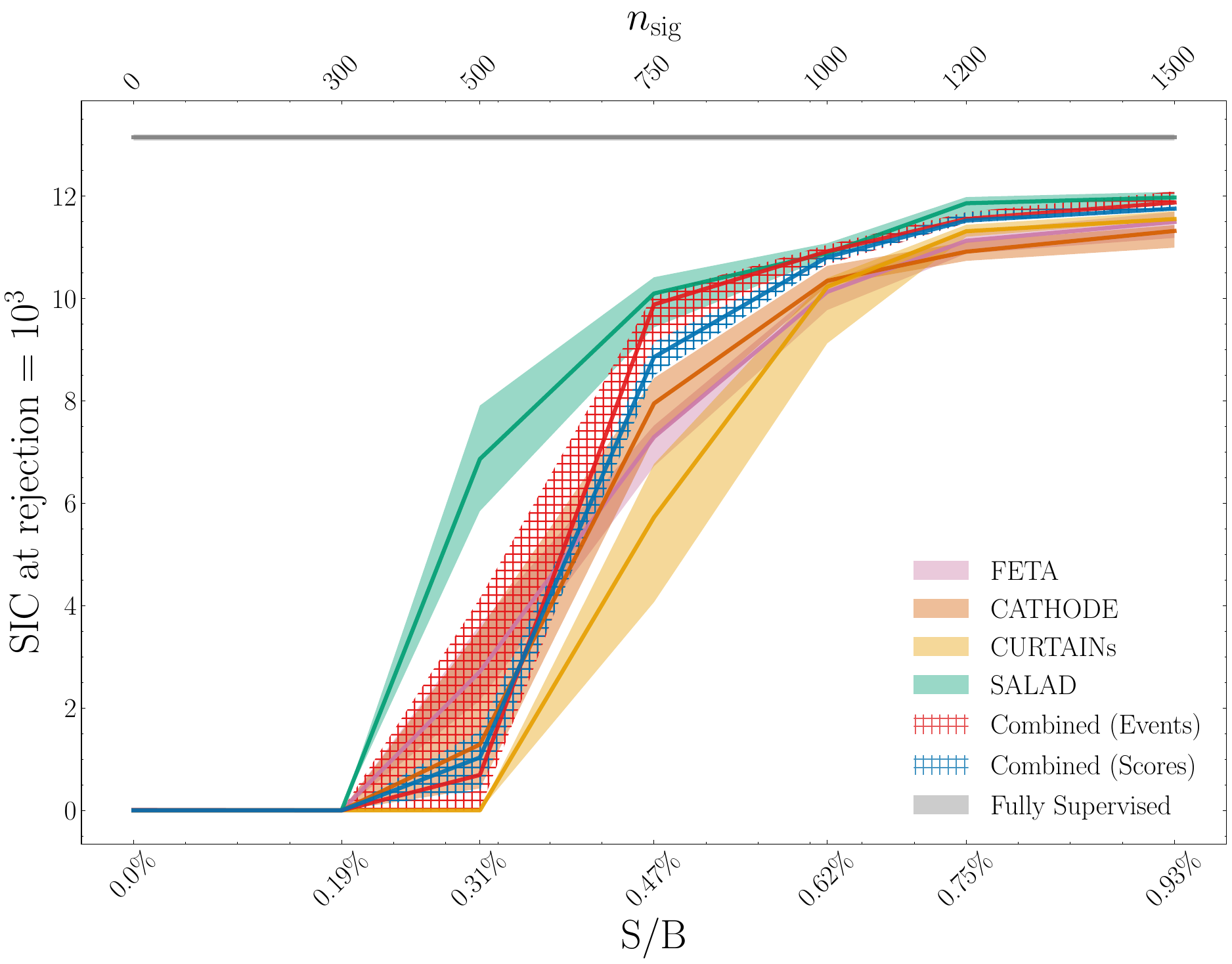}
   \caption{\label{fig:sic_at_rej_genseed2} SIC at a classifier rejection of $10^3$, all generator seeds set to ``2".}
  \end{subfigure}
  \\
    \begin{subfigure}[t]{.49\textwidth}
    \centering
    \includegraphics[width=\linewidth]{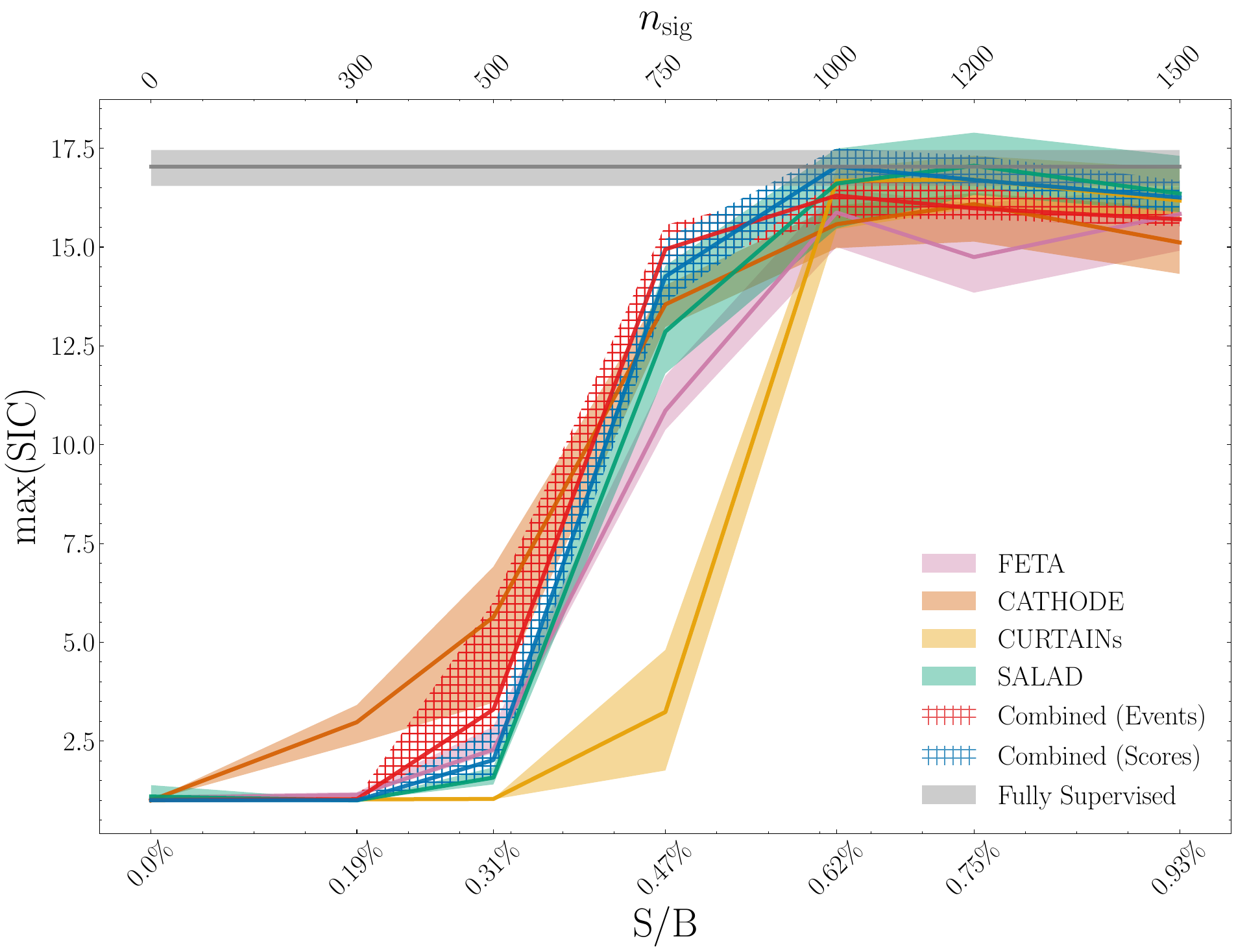}
    \caption{\label{fig:maxsic_genseed3}  Maximum of the SIC, all generator seeds set to ``3".}
  \end{subfigure}
  \begin{subfigure}[t]{.49\textwidth}
    \centering
    \includegraphics[width=\linewidth]{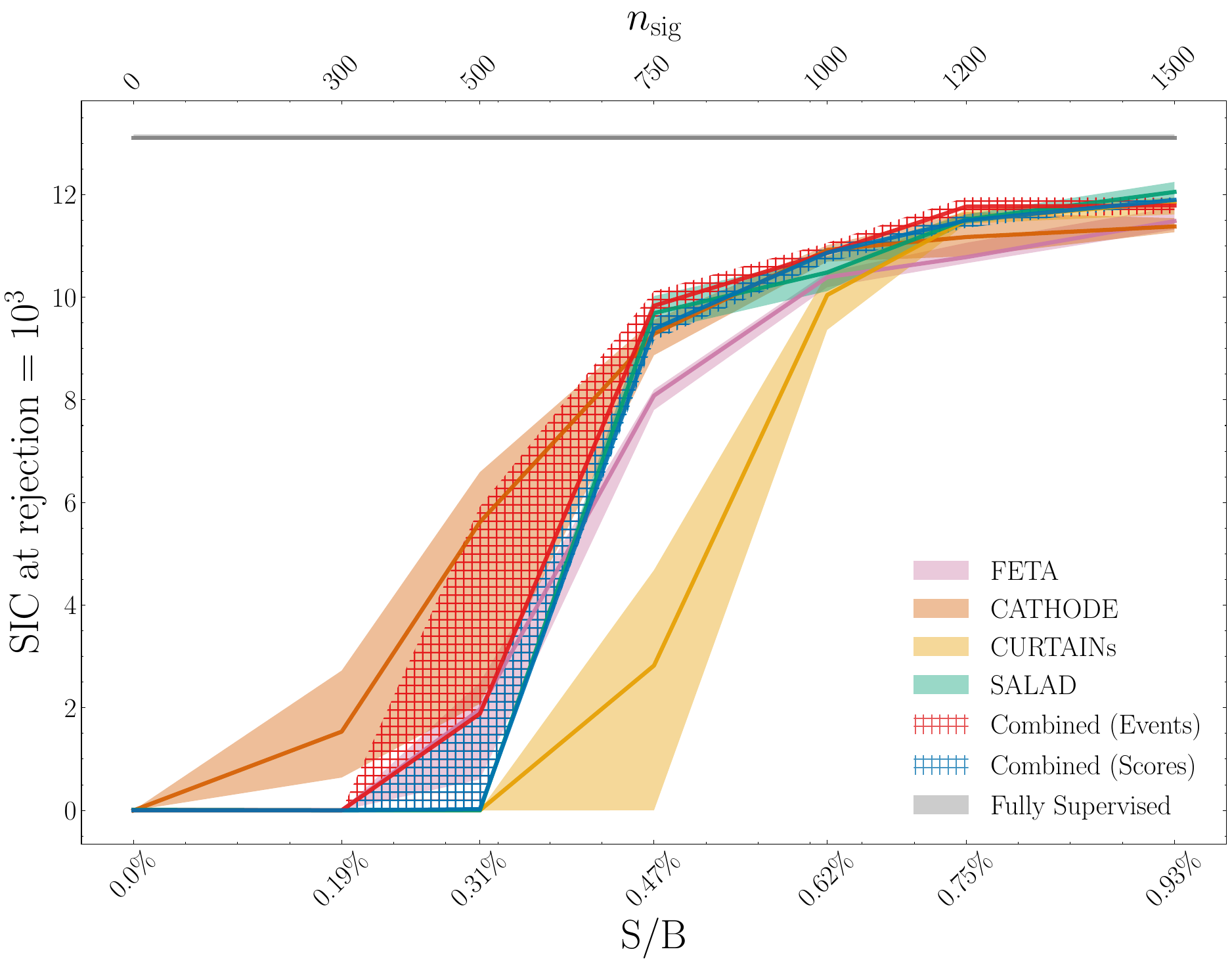}
   \caption{\label{fig:sic_at_rej_genseed3} SIC at a classifier rejection of $10^3$, all generator seeds set to ``3".}
  \end{subfigure}
  \\
    \begin{subfigure}[t]{.49\textwidth}
    \centering
    \includegraphics[width=\linewidth]{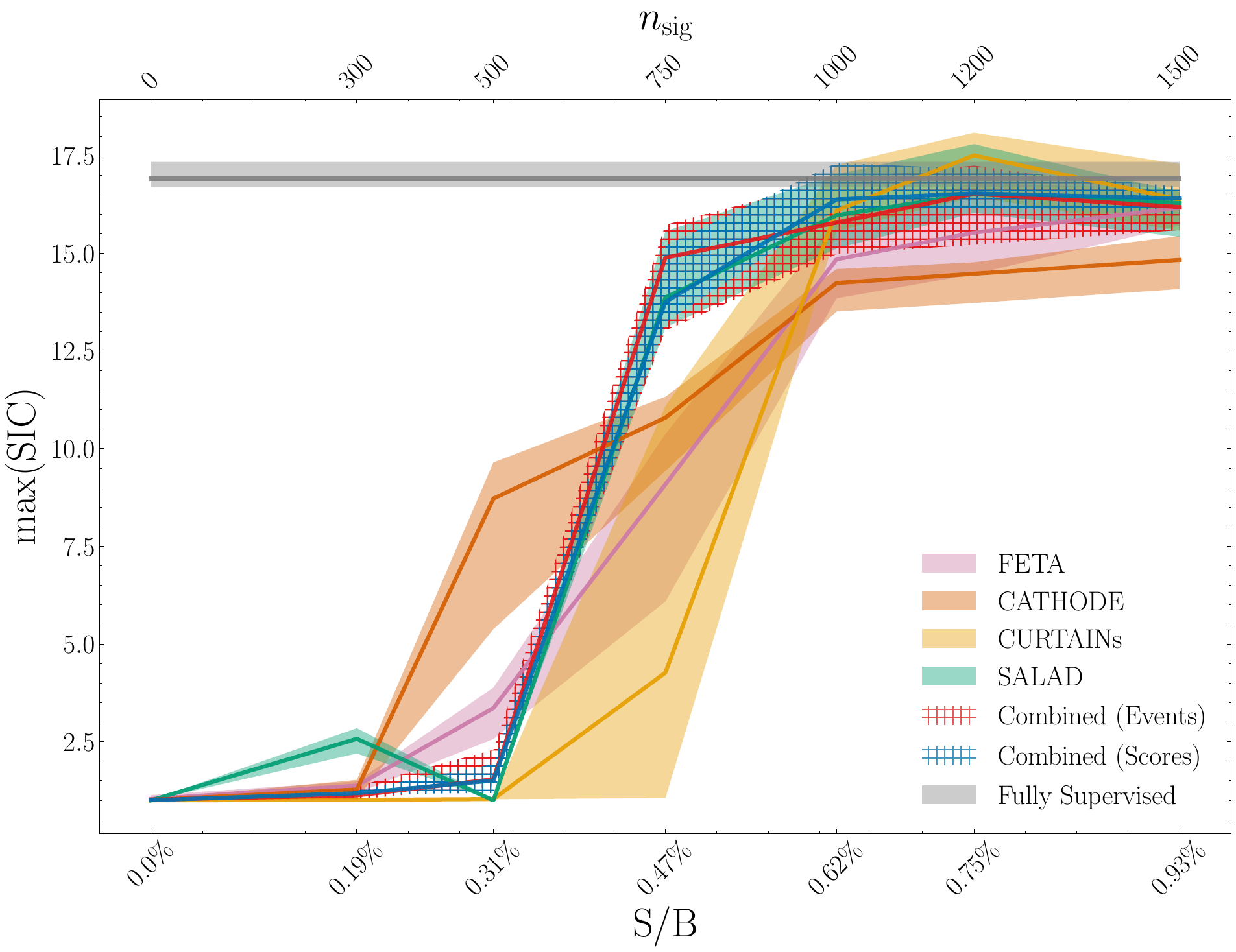}
    \caption{\label{fig:maxsic_genseed4}  Maximum of the SIC, all generator seeds set to ``4".}
  \end{subfigure}
  \begin{subfigure}[t]{.49\textwidth}
    \centering
    \includegraphics[width=\linewidth]{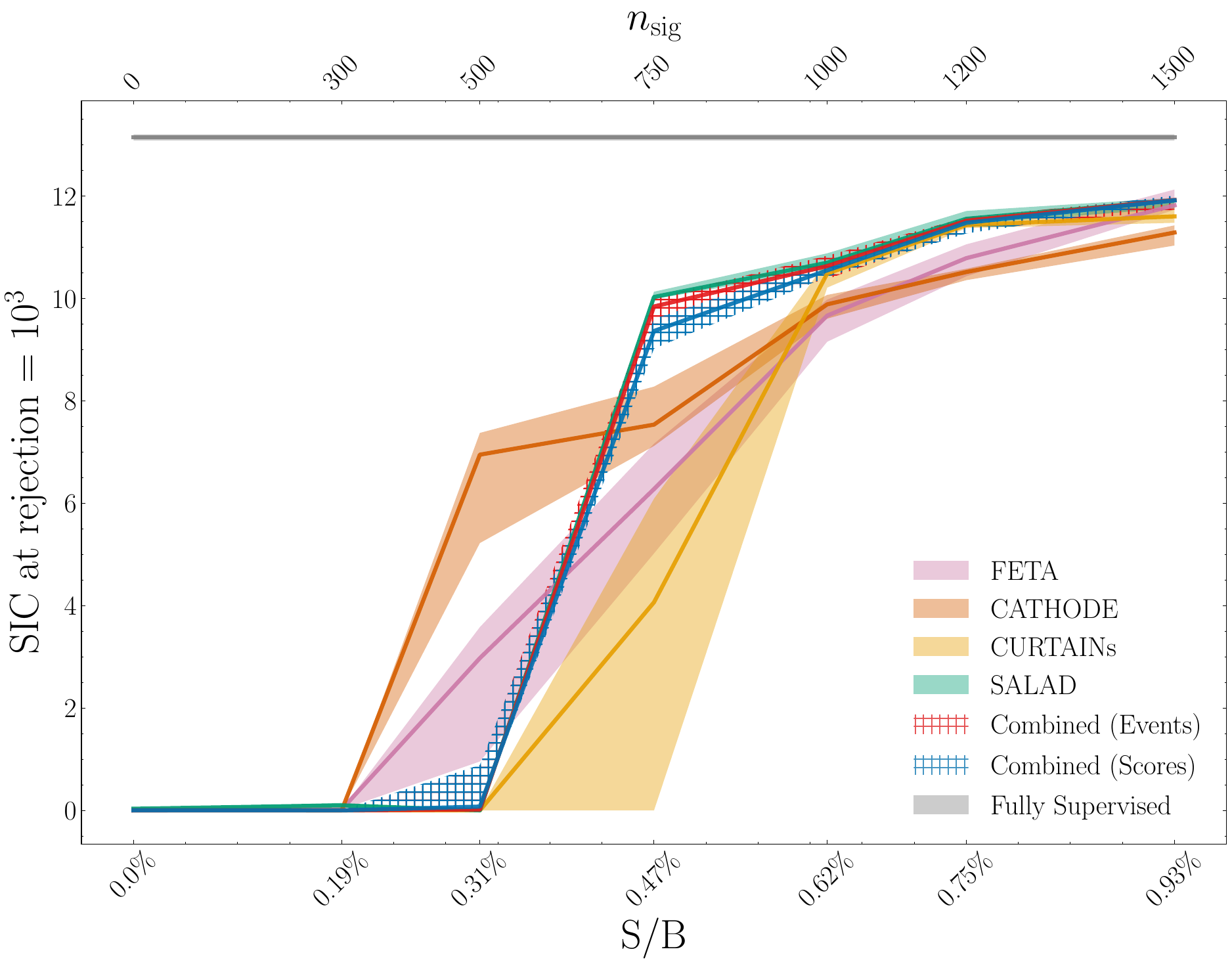}
   \caption{\label{fig:sic_at_rej_genseed4} SIC at a classifier rejection of $10^3$, all generator seeds set to ``4".}
  \end{subfigure}
\caption{\label{fig:bulk_metrics_single_genseed} Various metrics for a classifier trained to discriminate synthetic SM samples from data over a range of $n_\mathrm{sig}$ values. Equivalent to \Fig{fig:bulk_metrics}, but without averaging over generative network initializations.}
\end{figure}

While the previous section provides evidence in favor of robustness of network scores, there is a limit to this correlation with respect to the anomaly detection procedure. When the signal injection is below a certain threshold (which is likely to be model-dependent), the \textsc{CWoLa} procedure will break down, and even an idealized anomaly detection classifier will fail to pick up on the signal event. This can be seen in two ways:

\begin{enumerate}
    \item When the full sample combination study (corresponding to \Sec{sec:combining}) is rerun, regenerating all of the synthetic samples for each method by retraining the generative architectures with a different initialization, the performance of the individual methods is highly variable below $n_\textrm{sig} = 750$, but is stable above that point. In particular, if we do not average of generator initializations (as shown in the rows of \Fig{fig:bulk_metrics_single_genseed}), then for one instantiation, the \Salad{} method appears to win out at $n_\textrm{sig} = 500$; for another, the \Cathode{} method performs best at that signal injection; for a third, \Cathode{} and \Salad{} both exhibit low-$n_\mathrm{sig}$ fluctuations.
    
    \item From \Fig{fig:unification_percentile_5} (the fractional overlap between different sample generation methods of the 5th percentile of events classified as the most ``signal-like"), the agreement of the most anomalous signal events is relatively stable with respect to $n_\textrm{sig}$ above 750 events, but fluctuates highly below that value. This is again a sign that these is too little signal in the dataset for any of the individual learning methods to effectively recognize it. 
\end{enumerate}

\noindent Given these findings, another benefit of combination emerges: the combined methods break down below $n_\textrm{sig} = 750$, in a sense flagging the low signal statistics and indicating a poor regime to use \textsc{CWoLa} procedure. In this situaion, using an individual sample generation method might give an untrustworthy result. 

\clearpage

\section{Additional plots}
\label{app:more_plots}

In this section, we provide a small number of supplementary plots to complement the main text figures.

\begin{figure}[h]
    \centering
    \includegraphics[width = .8\textwidth]{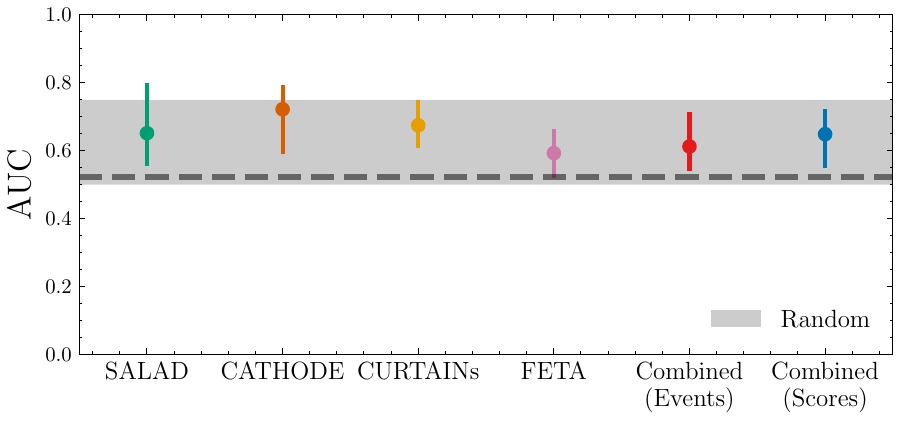}
    \caption{Receiver operating characteristic area-under-the-curves (AUC) for binary classifiers trained to discriminate each method's synthetic samples from data with $n_\mathrm{sig} = 0$. The table summarizes 100 classifier runs with different random seeds, with errorbars showing a 68-percentile spread.}
    \label{fig:aucs_w_combined}
\end{figure}

In \Fig{fig:aucs_w_combined}, we plot a companion plot to \Fig{fig:aucs}, illustrating the spread of ROC AUCs for a binary classifier trained to discriminate sets of synthetic samples (or their combination) from data with $n_\mathrm{sig} = 0$. Sample combination at both the event and the score level appears to more closely reproduce the spread coming from the random classifier, which indicates that the combination provides a more representative sample of SM background-like events.

\begin{figure}
    \centering
    \includegraphics[width=.6\linewidth]{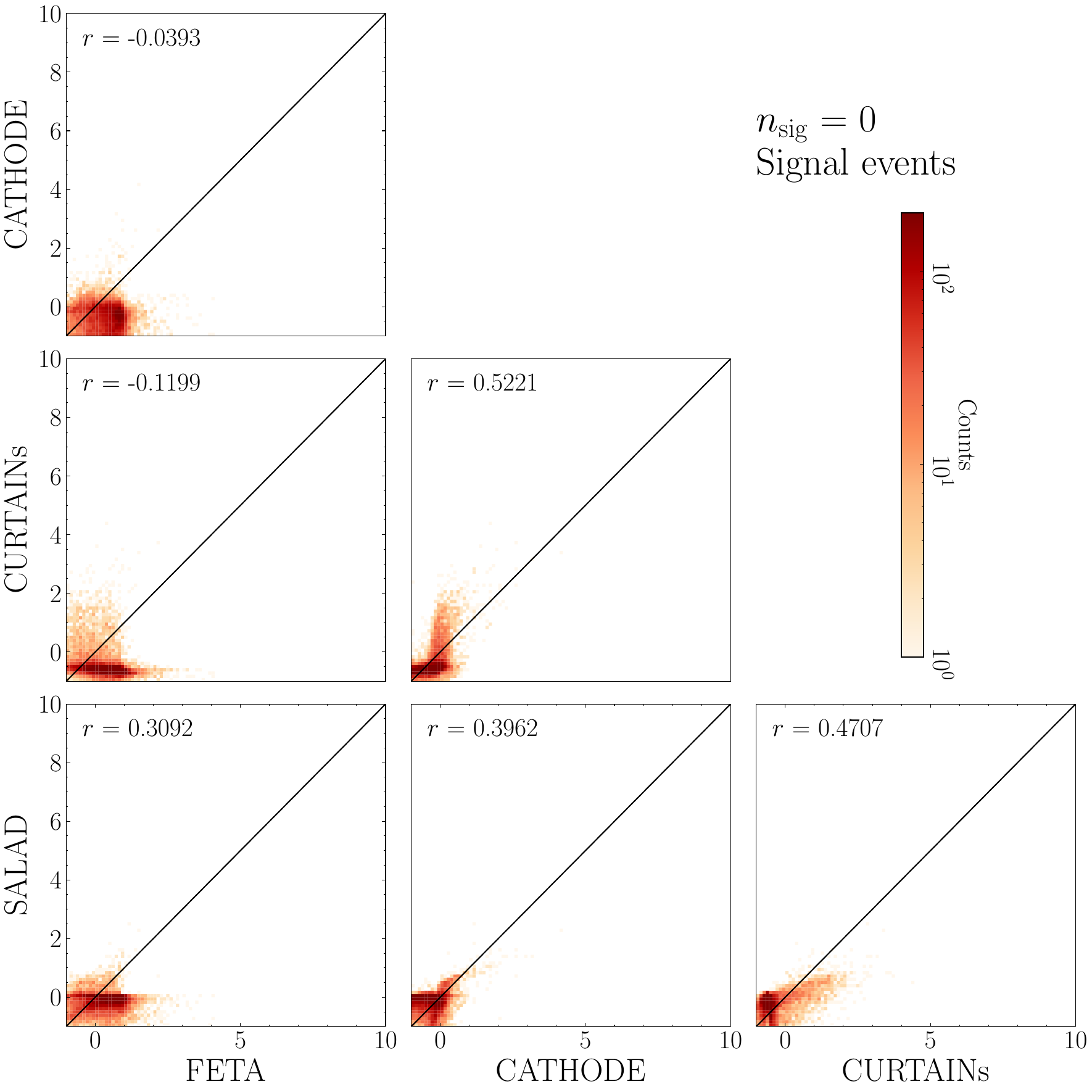}
    \caption{Scores for signal events for a binary classifier trained to discriminate synthetic SM background from data with $n_\mathrm{sig} = 0$). Each axis represents a different method of SM sample generation. $r$ denotes the Pearson correlation coefficient. The scores are standardized so as to make it clear what events the binary classifier flags as the most anomalous. For each method, scores are averaged over 10 classifier runs. This is a companion plot to \Fig{fig:corner_bkg_0_zoomed}}
    \label{fig:corner_sig_0}
\end{figure}

\begin{figure}
    \centering
    \includegraphics[width=.9\linewidth]{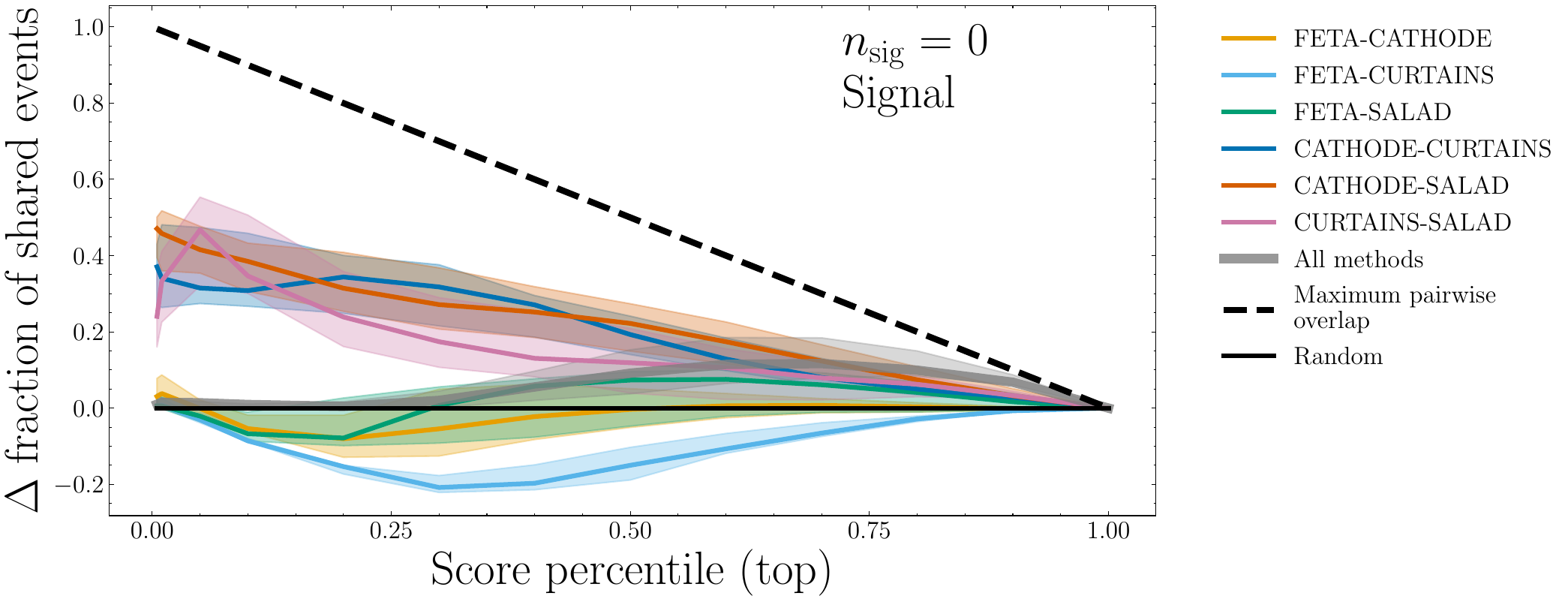}
    \caption{\label{fig:unification_0} Fractional overlap, with respect to a random-choice baseline, of the $p$th percentile of true signal events classified as the most ``signal-like" between different methods of synthetic SM sample generation. Errorbands show a 68-percentile spread across the median and come from 100 repetitions of training 5-fold classifiers on the associated methods with different random seeds and ensembling scores over 10 repetitions. This is a companion plot to \Fig{fig:unification_bkg_0}.}
\end{figure}

In \Fig{fig:corner_sig_0}, we plot the standardized scores for true signal events, as calculated by a binary classifier trained to discriminate the synthetic samples from data with $n_\mathrm{sig} = 0$. This is a companion plot to  \Fig{fig:corner_bkg_0_zoomed}. In \Fig{fig:unification_0}, we plot the fractional overlap, with respect to a random-choice baseline, of the $p$th percentile of true signal events classified as the most ``signal-like" between different methods of synthetic SM samples. This is a companion plot to  \Fig{fig:unification_bkg_0}.

\begin{figure}
\centering
  \begin{subfigure}[t]{.49\textwidth}
    \centering
    \includegraphics[width=\linewidth]{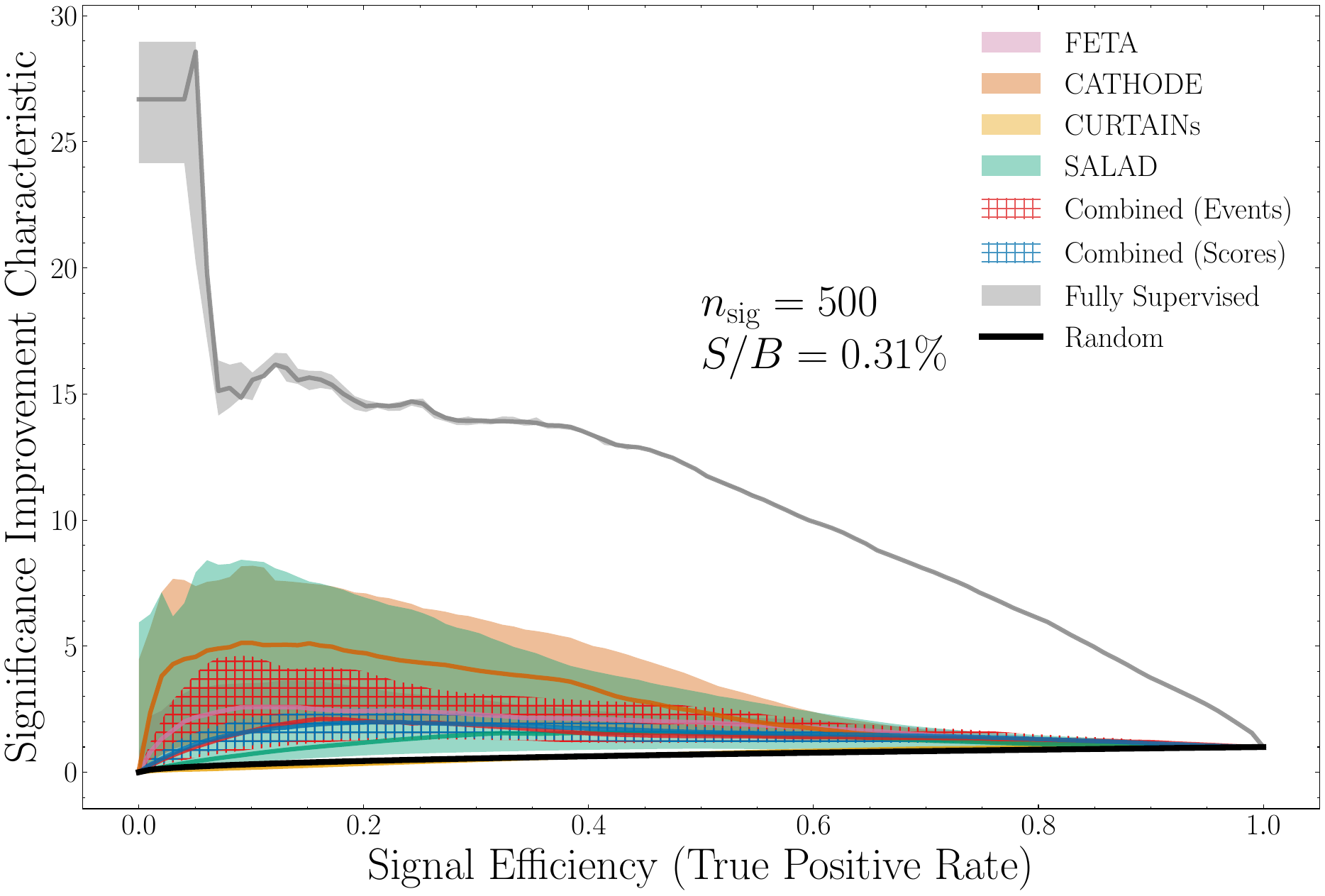}
    \caption{\label{fig:sic_allgen_500} Significance Improvement Characteristic plotted against the signal efficiency.}
  \end{subfigure}
  \begin{subfigure}[t]{.49\textwidth}
    \centering
    \includegraphics[width=\linewidth]{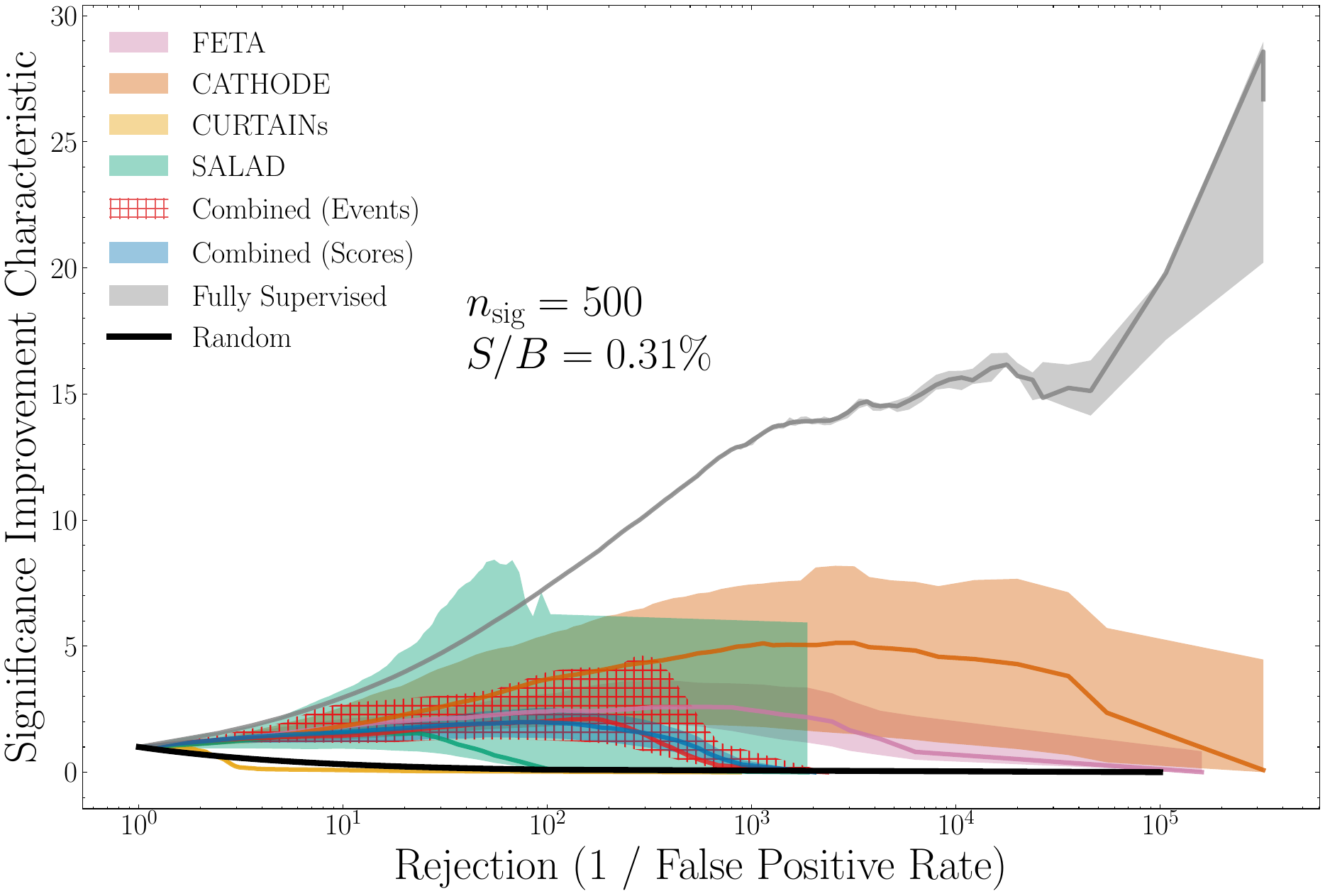}
   \caption{\label{fig:sic_vs_rej_allgen_500} Significance Improvement Characteristic plotted against classifier rejection.}
  \end{subfigure}
\caption{\label{fig:classifier_metrics_500} Various classifier metrics for a classifier trained to discriminate a combination of \Feta{}, \Cathode{}, and \Curtains{} synthetic SM samples from data with $n_\mathrm{sig} = 750$. Errorbands show
a 68-percentile spread across the median and come from training a 5-fold classifier 100 times with
different random seeds, over 3 independent generative model seeds, ensembling scores over 10 runs,
and averaging classifier metrics over the ensembles. This is a companion plot to \Fig{fig:classifier_metrics_750}.}
\end{figure}

In \Fig{fig:classifier_metrics_500}, we plot the Significance Improvement Characteristic, as a function of the signal efficiency and the rejection, for an ensemble of classifiers trained to discriminate a combination of \Feta{}, \Cathode{}, and \Curtains{} synthetic SM samples from data with $n_\mathrm{sig} = 500$. At this point, all methods fail to reliably pick up on this low signal injection. This is a companion plot to \Fig{fig:classifier_metrics_750}.

\end{document}